%% file: main.tex
\documentclass[letterpaper,twocolumn,10pt]{article}
\usepackage{usenix}

\usepackage{tikz}
\usepackage{amsmath}
\usepackage{comment}
\usepackage{amssymb}

\usepackage{filecontents}

\usepackage{amsmath,amsfonts}
\usepackage{multirow}
\usepackage{graphicx}
\usepackage{float}
\usepackage{url}

\usepackage{breakurl}
\usepackage{hyperref}
\usepackage{stfloats} 
\usepackage{makecell}
\usepackage{pifont}
\usepackage{threeparttable}
\usepackage{booktabs}
\usepackage{xspace}

\usepackage{algorithm}
\usepackage{algorithmicx}
\usepackage{algpseudocode}

\usepackage{subfigure}

\usepackage{listings}
\usepackage{xcolor}

\definecolor{codegreen}{rgb}{0,0.6,0}
\definecolor{codegray}{rgb}{0.5,0.5,0.5}
\definecolor{codepurple}{rgb}{0.58,0,0.82}
\definecolor{backcolour}{rgb}{0.97,0.97,0.97}

\lstdefinestyle{mystyle}{
    backgroundcolor=\color{backcolour},   
    commentstyle=\color{codegreen},
    keywordstyle=\color{magenta},
    numberstyle=\tiny\color{codegray},
    stringstyle=\color{codepurple},
    basicstyle=\ttfamily\scriptsize,
    breakatwhitespace=true,         
    breaklines=true,                 
    captionpos=b,                    
    keepspaces=true,   
    frame=lines,
    numbers=left,                    
    numbersep=5pt,                  
    showspaces=false,                
    showstringspaces=false,
    showtabs=false,                  
    tabsize=2
}
\usepackage[table]{xcolor}
\usepackage{array}
\newcommand{\PreserveBackslash}[1]{\let\temp=\\#1\let\\=\temp}
\newcolumntype{C}[1]{>{\PreserveBackslash\centering}p{#1}}
\newcolumntype{R}[1]{>{\PreserveBackslash\raggedleft}p{#1}}
\newcolumntype{L}[1]{>{\PreserveBackslash\raggedright}p{#1}}

\newcommand{\mname}{CuFuzz\xspace}

\newcommand{\etal}{\textit{et al.}}

\newcommand{\redcircle}[1]{%
  \tikz[baseline=(char.base)]{
    \node[shape=circle, draw=black, fill=red, text=white, inner sep=0.5pt, minimum size=1em] (char) {%
      \fontsize{0.7em}{0.7em}\selectfont #1
    };
  }%
}

\newcommand{\niparagraph}[1]{\noindent\textbf{#1}}

\begin{document}

\date{}

\title{\Large \bf CuFuzz: Hardening CUDA Programs through Transformation and Fuzzing}

\author{
{\rm Saurabh Singh}\\
Georgia Tech\\
saurabh.s@gatech.edu
\and
{\rm Ruobing Han}\\
Georgia Tech\\
rhan38@gatech.edu
\and
{\rm Jaewon Lee}\\
Georgia Tech\\
jaewon.lee@gatech.edu
\and
{\rm Seonjin Na}\\
Georgia Tech\\
seonjin.na@gatech.edu
\and
{\rm Yonghae Kim}\\
Intel Labs\\
yonghae.kim@intel.com
\and
{\rm Taesoo Kim}\\
Georgia Tech\\
taesoo@gatech.edu
\and
{\rm Hyesoon Kim}\\
Georgia Tech\\
hyesoon@cc.gatech.edu
} 

\maketitle

\input{sources/00_abstract}

\input{sources/01_introduction}

\input{sources/02_background}
\input{sources/04_errors}

\input{sources/05_GPUFuzz}

\input{sources/06_optimization}

\input{sources/08_gms}

\input{sources/07_eval}

\input{sources/09_related}

\input{sources/10_conclusion}

\cleardoublepage
\appendix

\cleardoublepage
\bibliographystyle{plain}
\bibliography{sources/references}

\end{document}

%% file: sources/00_abstract.tex
\begin{abstract}

GPUs have gained significant popularity over the past decade, extending beyond their original role in graphics rendering. This evolution has brought GPU security and reliability to the forefront of concerns. Prior research has shown that CUDA’s lack of memory safety can lead to serious vulnerabilities. While fuzzing is effective for finding such bugs on CPUs, equivalent tools for GPUs are lacking due to architectural differences and lack of built-in error detection.

In this paper, we propose \mname, a novel compiler-runtime co-design solution to extend state-of-the-art CPU fuzzing tools to GPU programs. \mname transforms GPU programs into CPU programs using compiler IR-level transformations to enable effective fuzz testing. To the best of our knowledge, \mname is the first mechanism to bring fuzzing support to CUDA, addressing a critical gap in GPU security research. By leveraging CPU memory error detectors such as Address Sanitizer, \mname aims to uncover memory safety bugs and related correctness vulnerabilities in CUDA code, enhancing the security and reliability of GPU-accelerated applications. To ensure high fuzzing throughput, we introduce two compiler-runtime co-optimizations tailored for GPU code: Partial Representative Execution (PREX) and Access-Index Preserving Pruning (AXIPrune), achieving average throughput improvements of 32× with PREX and an additional 33\% gain with AXIPrune on top of PREX-optimized code. Together, these optimizations can yield up to a 224.31x speedup. In our fuzzing campaigns, \mname uncovered 122 security vulnerabilities in widely used benchmarks.

\end{abstract}

%% file: sources/01_introduction.tex
\section{Introduction}
\label{sec:intro}

Ensuring memory safety is one of the biggest challenges in software security, especially in languages like C and C++ that offer low-level memory control with limited safety checks. Memory safety violations such as buffer overflows and out-of-bounds accesses account for 60--70\% of vulnerabilities in iOS and macOS~\cite{apple_mem_safety}, 70\% of vulnerabilities in Microsoft products~\cite{microsoft_mem_safety}, and 90\% of vulnerabilities in Android~\cite{google_mem_safety}. Google reports that 75\% of zero-day CVEs exploited in the wild involve memory safety issues~\cite{google_zeroday_report}. Recognizing the severity of this problem, the White House ONCD recently issued a national call to action, urging a shift to memory-safe languages to eliminate classes of memory-related vulnerabilities~\cite{oncd24_whitehouse_memsafety_report}.

Today, GPUs are widely used to accelerate applications like machine learning~\cite{chetlur2014cudnn}, weather prediction~\cite{gpu_appl_weatherpred}, scientific simulations~\cite{gpu_appl_scientificsim}, cryptography~\cite{wang2012accelerating,manavski2007cuda}, autonomous vehicle navigation~\cite{gpu_appl_autonav}, etc. GPU architectures have traditionally prioritized performance, throughput, and parallelism, with security receiving comparatively little attention. As GPUs are increasingly adopted for applications that manage critical and sensitive data, ensuring the security of both GPU hardware and software is becoming an urgent priority. Recent reports show that attackers have started exploiting coprocessor vulnerabilities, including those in GPUs, to bypass system-level mitigations, a trend that is likely to continue due to limited protections in such devices~\cite{google_mte_coprocessor_vuln}.

CUDA is the dominant programming language for GPUs built on top of C/C++. It inherits many advantages of C/C++, such as high speed, simplicity, and low-level control, making it highly effective for performance-intensive applications. However, it also inherits the lack of memory safety in C/C++, which opens up the potential for memory safety vulnerabilities in CUDA code~\cite{corr2015_bufovf_cuda, springer2016_ovf_vuln_in_gpus}. Recent studies have demonstrated that GPU memory safety vulnerabilities can be exploited to craft sophisticated attacks, such as Mind Control~\cite{mind_contol_attack_2021} and Return-Oriented Programming (ROP)~\cite{fun_and_profit_2024}. 

Fuzzing~\cite{afl, libfuzzer, radamsa, honggfuzz} has long been utilized as an alternative to the daunting task of identifying bugs and vulnerabilities manually. Fuzzing aims to trigger crashes and other unintended issues in a program by feeding it with unexpected and randomly generated inputs. This approach is especially effective at uncovering memory-related issues, such as buffer overflows and memory corruption, and can expose vulnerabilities in complex systems with unpredictable runtime behavior. Fuzzing’s ability to generate large numbers of inputs to exhaustively exercise control flow paths in a program makes it highly effective for uncovering security vulnerabilities. 

While fuzzing has been employed in the CPU domain for decades with tools such as AFL and ASan, the GPU fuzzing infrastructure is still in its infancy. Extending existing fuzzing techniques to GPUs presents several unique challenges due to the architectural and programming differences between CPUs and GPUs. GPUs are designed for high-throughput, allowing them to handle thousands of threads concurrently. This parallelism, while advantageous for performance, creates significant complexity for fuzzing. Additionally, GPUs use a complex memory hierarchy comprising several memory spaces, such as global, local, shared, constant, and texture memories. Most importantly, GPUs do not support precise exceptions, making it inherently difficult to trigger an error or generate a crash, which is the primary feedback mechanism in CPU fuzzing. This intricate memory structure, combined with a lack of hardware protection mechanisms, makes it challenging to develop fuzzers for GPU programs.

To address these challenges, we propose \mname, a compiler–runtime co-design framework that integrates existing CPU fuzzing infrastructure with GPU programs. \mname translates CUDA code into CPU code while preserving the original program’s memory behavior, enabling the direct use of tools such as AFL++ and AddressSanitizer. This provides an end-to-end fuzzing framework and, by bridging this gap, \mname delivers the first practical and scalable solution for fuzzing GPU programs.

\mname's translated fuzzing approach is attractive for several reasons. First, it allows reuse of existing CPU fuzzing infrastructure, including mature tools like Address Sanitizer and Valgrind, as well as widely available CPU servers and data centers. This eliminates the need for setting up dedicated GPU servers, reducing costs and the barrier to entry. Second, CPUs provide better support for runtime error detection, whereas GPUs lack precise exceptions and often delay error reporting. Third, translated fuzzing enables testing GPU programs across diverse CPU architectures with unique safety features such as MTE on ARM~\cite{ref_mte}, MPX on Intel~\cite{intex_mpx}, and PMP on RISC-V~\cite{riscv_priv_spec} CPUs, thereby improving bug detection. Finally, launching a kernel on GPUs involves setup and excessive data transfer between host and device. In contrast, translated programs avoid this setup cost, leading to faster fuzzing iterations and efficient bug discovery. Together, these advantages make translated fuzzing a practical and scalable strategy for uncovering GPU memory safety issues.

However, enabling high-throughput fuzzing requires more than just translation. Naïvely translating GPU code to CPU code can introduce significant performance overhead due to the mismatch in execution models. This is particularly problematic for fuzzing, which requires executing the program thousands of times to explore all control-flow paths. To reduce this overhead, we introduce two compiler-runtime optimizations specifically tailored for fuzzing GPU code: 1) PREX (Partial Representative Execution) and AXIPrune (Access-Index Preserving Pruning). These optimizations achieve a speedup of up to 224.31x (average: 32.73x) over naively translated CUDA code. \mname found 122 unique issues during fuzzing campaigns that include 15 kernel crashes, 40 host crashes, and 67 hangs.

Overall, this paper offers several noteworthy contributions:
\begin{itemize}
    \item We present \textbf{\mname}, a novel end-to-end compiler-runtime co-design framework that enables effective fuzzing of CUDA programs by transforming GPU programs into CPU programs, unlocking the use of mature CPU fuzzing infrastructure directly on GPU programs.

    \item We introduce two novel compiler-runtime optimizations to improve fuzzing throughput, reduce redundant computation, and accelerate bug discovery.
   
\end{itemize}

%% file: sources/02_background.tex
\section{Background}
\label{sec:background}

\subsection{Software Fuzzing}

Fuzzing is an automated software testing technique that discovers vulnerabilities by feeding a program systematically generated inputs. It identifies memory bugs like buffer overflows, out-of-bound reads/writes, and segmentation faults. Fuzzing is effective for finding security vulnerabilities that manual testing or static analysis might miss. There are various types of fuzzing: black-box, white-box, and gray-box. Black-box fuzzers test without knowledge of the program's internal structure, while white-box fuzzers use the program's source code to generate inputs. Gray-box fuzzers use limited runtime information, with coverage-guided fuzzers like AFL/AFL++~\cite{afl,aflpp} using runtime coverage data to maximize input generation.

\begin{figure}[h]
    \centering
    \includegraphics[width=1.0\linewidth]{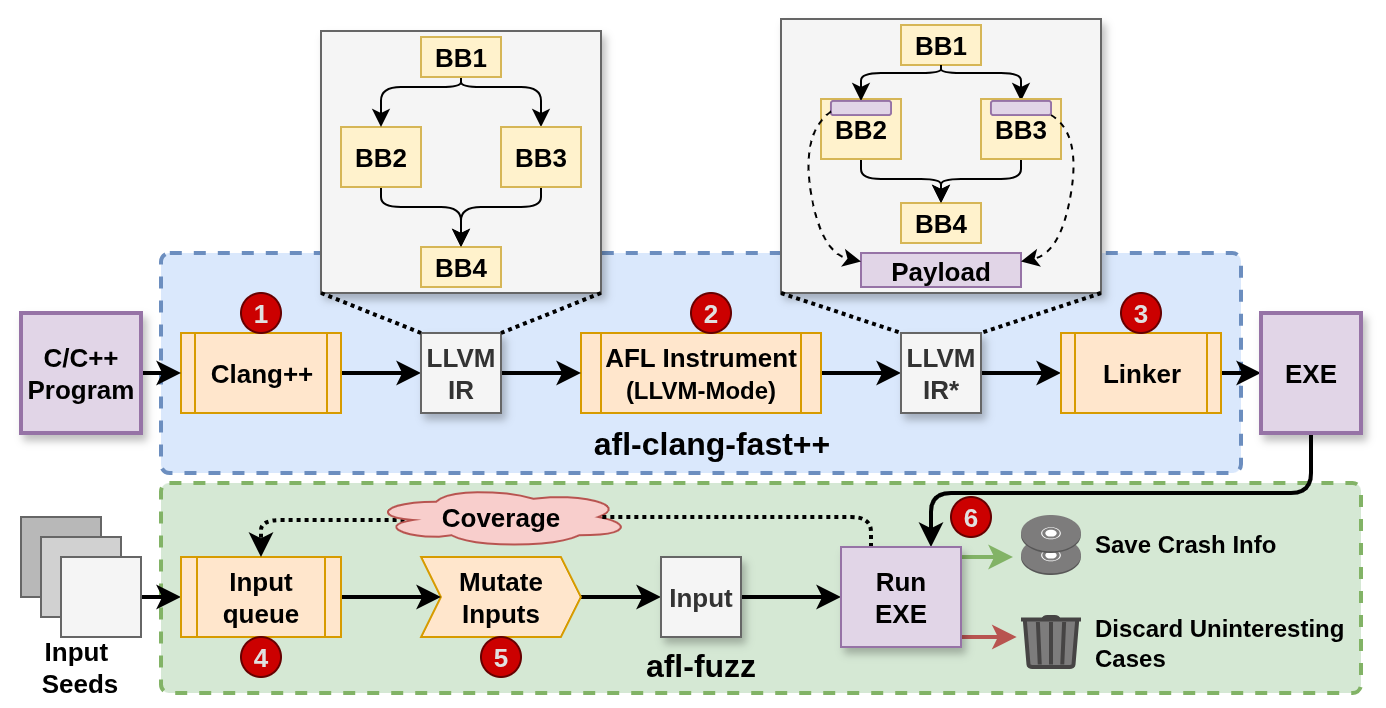}
    \caption{AFL LLVM-Mode Workflow.}
    \label{fig:afl_workflow}
\end{figure}

AFL mainly consists of two parts: a compiler wrapper, which acts as a drop-in replacement for existing compilers such as Clang \& GCC and is used to instrument the code, and a coverage-guided fuzzing driver (\textit{afl-fuzz}). The detailed workflow of AFL++ in LLVM instrumentation mode is shown in Figure~\ref{fig:afl_workflow}.
When the AFL compiler is invoked~\redcircle{1}, internally, it generates LLVM IR corresponding to the source code and uses AFL LLVM Pass to instrument it~\redcircle{2}. The instrumentation consists of 1) a payload code that keeps track of coverage data during run-time and 2) trampoline jumps to payload code that are injected after branch instructions in the IR. In the final step, the linker produces an executable. For fuzzing, the user must provide a set of seed inputs for AFL++.

During fuzzing, \textit{afl-fuzz} runs the test program as a child process, providing inputs and extracting coverage data~\redcircle{6}. AFL uses a genetic algorithm to generate the next generation of inputs based on random mutations of the previous generation of inputs~\redcircle{5}. Input traits from previous generations are passed to the next generation based on their fitness. Fitness, in this context, can be defined as a measure of an input's ability to discover new control flow paths during the code execution. Consequently, inputs that do not discover new control flow paths are discarded. In this way, AFL tries to find all possible control flow paths that can be executed. Some control flow paths result in crashes or hangs, in which case AFL saves the corresponding inputs for the user to analyze later.

Fuzzers like AFL++ are often used with tools like AddressSanitizer (ASan)~\cite{asan_2012} to improve the detection of memory errors during fuzzing. ASan is a runtime memory error detector that identifies issues like buffer overflows, use-after-free errors, and stack overflows. When used together, AFL++ instruments the code to gather coverage data, while ASan monitors for memory violations that may be triggered by the generated inputs. A fuzzer is as good as the underlying error detector. Without an effective detector like ASan, AFL++ might miss crucial vulnerabilities, as it would lack the capability to uncover subtle memory issues that can lead to security risks in production code.

\subsection{Memory Safety on CPUs vs GPUs}
\label{sec:memsafety_cpu_vs_gpu}

Memory safety errors can be broadly divided into two categories: Spatial and Temporal. Spatial memory errors arise when a program tries to access data outside its intended bounds. Temporal memory errors arise when a program tries to access the data beyond its life cycle.

On CPUs, memory safety errors fall into 4 categories: 1) \textit{Buffer overflows}: linear accesses beyond a buffer’s boundary, 2) \textit{Out-of-bounds (OOB) read/write}: arbitrary accesses outside the allocated regions, 3) \textit{Dangling pointers}: Accessing freed memory region, and 4) \textit{Uninitialized memory accesses}. In CPUs, tools like Address Sanitizer (ASan)~\cite{asan_2012}, Memory Sanitizer (MSan)~\cite{msan_2015}, and Valgrind~\cite{valgrind_2007} help detect these bugs. Additionally, CPUs have built-in hardware and OS support (e.g. segmentation faults) that help detect such violations. 

On GPUs, these errors arise in a more complex environment due to the highly parallel nature of execution. Out-of-bounds accesses in GPUs can occur due to threads trying to access memory outside the assigned regions. Without robust hardware-level error reporting, these errors often go undetected. A recent study demonstrated that OOB errors can occur in all three GPU memory spaces (global, local, and shared). Furthermore, 
return address corruption-based control flow hijacking attacks are also possible on modern GPUs~\cite{fun_and_profit_2024}. In our experiments we found that enabling device debug symbols (\texttt{-g -G}) often exposes additional vulnerabilities.

%% file: sources/04_errors.tex
\subsection{GPU Programs}
GPU programs are divided into host and kernel code. 
The host code is compiled using the standard CPU compiler (e.g. \texttt{gcc} or \texttt{clang}), while the kernel code is lowered to NVIDIA SASS assembly using \texttt{nvcc}. After that, both are linked with the C/C++ and CUDA runtimes and packaged into a fat binary with entry point residing in the host code.

\subsubsection{GPU Host Code Errors}  
Host code runs on the CPU and manages GPU resources, including memory allocation, copying data from host to GPU memory, and kernel launches. It is written in C++ with additional syntax to support GPU operations. Therefore, all the errors discussed in section~\ref{sec:memsafety_cpu_vs_gpu} are applicable to the host code. Additionally, host code is also prone to GPU specific errors such as: 1) GPU memory allocation failure (\texttt{cudaMalloc}) when device memory is insufficient to allocate the requested buffers, 2) Invalid kernel launch parameters such as wrong grid or block dimensions (reported as \texttt{cudaErrorInvalidConfiguration}), and 3) Misuse of the CUDA API, such as failing to check error codes or using APIs in improper states. Further, improper resource management such as failing to release memory or terminate streams and events can result in memory leaks and resource exhaustion.

\subsubsection{GPU Kernel Code Errors}
Kernel code errors occur during device execution and are particularly harder to debug due to massive parallelism and limited hardware support for detection. A common error is out-of-bounds memory access, where threads attempt to read/write memory outside the allocated ranges in global, local, or shared memory. Apart from that, errors like use-after-scope (UAS), use-after-free (UAF), and invalid free (IF) are also possible if the device heap is used. Additionally, errors like divide-by-zero, floating-point error, etc. can also occur.

\section{Threat Model}
\label{sec:threat_model}

We consider GPU-accelerated applications that process untrusted inputs via CUDA kernels (e.g., LLM inference, analytics). We assume the host OS, CUDA runtime/driver, and hardware are trusted, with no physical attacker access. Our threat model targets \textbf{end-user to service attacks}, where crafted inputs reach vulnerable kernels, potentially causing GPU memory corruption, control-flow hijacking, data leakage, or service crashes, analogous to classical CPU-side exploits. We focus on memory-safety violations in both CUDA kernels and host code, and their impact on the confidentiality, integrity, and availability of GPU-accelerated systems.

%% file: sources/05_GPUFuzz.tex
\section{Overview of \mname}
\label{sec:gpufuzz_approach}

\begin{figure*}[tbp]
    \centering
    \includegraphics[width=0.9\textwidth]{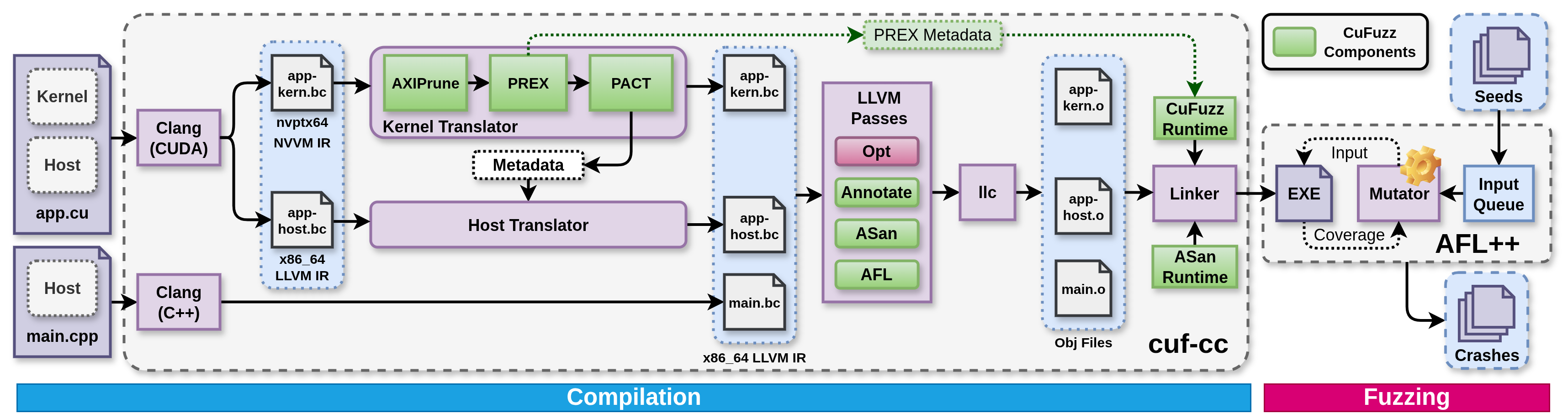}
    \caption{End-to-end \mname workflow: Translating GPU programs via \texttt{cuf-cc} and fuzzing them with AFL++}
    \label{fig:gpufuzz_workflow}
\end{figure*}

\mname workflow consists of two stages: compilation and fuzzing. Compilation is handled by \texttt{cuf-cc}, a drop-in replacement for \texttt{nvcc}/\texttt{clang} that encapsulates translation and instrumentation, simplifying fuzzing setup and integration with build tools. As shown in Figure~\ref{fig:gpufuzz_workflow}, \texttt{cuf-cc} uses \texttt{clang++} to compile CUDA code into two intermediate representations (IR): 1) NVVM IR (target triple: \textit{nvptx64}) for the kernel code and 2) LLVM IR (target triple: \textit{x86\_64}) for the host code. 

The \textbf{kernel translator} transforms the NVVM IR into LLVM IR suitable for CPU execution. During this transformation, a sequence of analysis and optimization passes are applied. First, the AXIPrune pass prunes expensive computation operations that do not influence the kernel’s memory access index, simplifying the kernel logic while preserving memory access behavior. Next, the PREX analysis pass examines the memory access patterns in the kernel and determines whether they are affine with respect to thread and block indices. Finally, this analysis guides the PACT transformation pass, which flattens the GPU thread hierarchy and maps them to CPU threads and loops. PACT also generates metadata about the kernel, which is passed to the host translator. Section~\ref{sec:optimizations} describes AXIPrune and PREX and Section~\ref{sec:PACT} presents the PACT transformation.

The \textbf{host translator} transforms host-side LLVM IR by removing CUDA-specific constructs or replacing them with CPU equivalents. CUDA built-in abstractions such as accesses to constant memory are substituted with normal memory operations in the CPU heap. It also replaces CUDA kernel launch syntax into direct calls to the translated kernel function with appropriate argument packing. Additionally, it removes CUDA runtime calls such as \texttt{cudaDeviceSynchronize()}, which are no longer necessary in the CPU-execution context. 

After translation, all LLVM IRs target the host triple. Next, a sequence of passes are applied to optimize and instrument the code for fuzzing, beginning with \texttt{-O2} optimization to improve efficiency. When compiling with \texttt{-fsanitize=address}, functions are annotated with the \texttt{SanitizeAddress} attribute, allowing ASan to insert memory-safety checks. However, in \mname workflow, the kernel IR is derived from CUDA code and lacks these annotations, leaving kernel functions uninstrumented. To address this, \texttt{cuf-cc} utilizes a custom pass that annotates all kernel functions, ensuring they are subsequently instrumented by ASan pass for memory-safety checks.

Finally, the AFL~\cite{aflpp} instruments the IR by adding payload code and injecting trampoline jumps after branch instructions, allowing tracking of branch-level coverage at runtime. During fuzzing, AFL fuzzer can use this coverage data to guide input mutation and exploration of new execution paths. 

After instrumentation, the IR is compiled with \texttt{llc} and linked with \mname's runtime libraries to produce the final executable. These runtime libraries (1) provide CPU-side implementations of common CUDA functions (e.g., \texttt{cudaMalloc()}, \texttt{cudaMemcpy()}), enabling the translated program to run independently, and (2) implement a thread-pool abstraction where each CPU thread executes multiple GPU thread blocks sequentially, enabling parallel execution across multiple threads. To further improve coverage, the runtime libraries are also instrumented with ASan, allowing memory safety violations within the emulated CUDA functions or kernel execution to be detected during fuzzing.

\subsection{PREX-Aware Collapsing Transform}
\label{sec:PACT}
The \mname kernel translator maps the GPU thread hierarchy onto the CPU execution model at the LLVM IR level using PREX-Aware Collapsing Transform (PACT). The key idea of GPU-to-CPU migration is to bridge the gap in the degree of parallelism between GPUs and CPUs. GPUs typically launch a large number of threads, whereas CPUs support only a limited number of concurrent threads. To bridge this gap, PACT transform maps a GPU block to a single CPU thread, similar to prior work~\cite{manavski2007cuda,cox,cupbop_todaes}. The threads within a block are mapped to iterations of a for-loop in the transformed CPU code. This is illustrated in Listings~\ref{lst:pact_orig_gpu_prog} and~\ref{lst:pact_transformed_gpu_prog}, which show a simple vector addition kernel and its corresponding CPU version after transformation respectively. While the actual transformation takes place at the LLVM IR level, we present simplified source code for the ease of understanding. 

As shown in Listing~\ref{lst:pact_orig_gpu_prog}, the original GPU program requires \( \texttt{gridSize} \times \texttt{blockSize} \) threads (line 15), whereas the transformed CPU program (Listing~\ref{lst:pact_transformed_gpu_prog}), only requires \( \texttt{gridSize} \) CPU threads(line 17). In the transformed program, each CPU thread executes a function that wraps the GPU kernel with for-loops (lines 6 and 12) iterating \( \texttt{blockSize} \) times. Consequently, each GPU thread is mapped to a single iteration of for-loop in the transformed CPU program. To preserve the barrier semantics (line 10), PACT splits the kernel into two loops before and after the barrier. Since local variable \( id \) is used on both sides of the barrier, it is converted into an array (line 5) to maintain correctness across loop boundaries. As a result, barriers in the GPU program introduce additional overheads in the transformed CPU programs. We introduce an optimization to reduce the overhead by barriers and other similar instructions in Section~\ref{sec:aggressive_pruning}. 

Beyond bridging the gap in parallelism, GPU-to-CPU migration also remaps the GPU memory hierarchy to the CPU memory hierarchy. Specifically, CUDA global memory is mapped to CPU heap, while CUDA shared memory and local memory are mapped to CPU stack. For instance, the two GPU shared memory variables (Listing~\ref{lst:pact_orig_gpu_prog}, lines 4 and 5) are transformed into stack variables in the CPU code (Listing~\ref{lst:pact_transformed_gpu_prog}, lines 2 and 3). This mapping preserves memory-safety bugs as buffer overflows in GPU global memory translate to heap overflows on the CPU, and violations in shared or local memory translate to stack errors after PACT transformation.

PACT also uses metadata from the PREX analysis which describes whether memory accesses in the kernel are affine with respect to thread and block indices. If all accesses are affine, PACT generates a simpler transformation where the thread index (\texttt{tid}) is passed as a function argument instead of using a loop. This reduces overhead and makes the transformed code more efficient. The host code is also updated to use this metadata correctly when calling the kernel. Details of PREX optimization are described in Section~\ref{sec:PREX}

\begin{lstlisting}[language=C++, caption=GPU Vector Addition Program, label={lst:pact_orig_gpu_prog}]
// Device code
__global__ void vecAdd(int *a, int *b, int *c) {
  // Shared Variables
  __shared___ s_a[blockSize];
  __shared___ s_b[blockSize];
  // Local Variables
  int id = blockIdx.x * blockSize + threadIdx.x;
  s_a[threadIdx.x] = a[id];
  s_b[threadIdx.x] = b[id];
  __syncthreads();
  c[id] = s_a[threadIdx.x] + s_b[threadIdx.x];
}
// Host code
int main() {
  vecAdd<<<gridSize, blockSize>>>(d_a, d_b, d_c);
}
\end{lstlisting}

\begin{lstlisting}[language=C++, caption=Transformed CPU Vector Addition Program,label={lst:pact_transformed_gpu_prog}]
void vecAdd(int *a, int *b, int *c, int bid) {
  int s_a[blockSize];
  int s_b[blockSize];
  // Scalar variable is translated to an array
  int id [blockSize];
  for(int tid = 0; tid < blockSize; tid++) {
    id[tid] = bid * blockSize + tid;
    s_a[tid] = a[id[tid]];
    s_b[tid] = b[id[tid]];
  }
  // __syncthreads()
  for(int tid = 0; tid < blockSize; tid++)
    c[id[tid]] = s_a[tid] + s_b[tid];
}
int main() {
  #pragma omp parallel for
  for(int bid = 0; bid < gridSize; bid++)
    vecAdd(d_a, d_b, d_c, bid);
}
\end{lstlisting}

{\bf Correctness of Transformation:} In the GPU \texttt{vecAdd}, each thread loads elements from global memory into shared memory arrays \texttt{s\_a} and \texttt{s\_b}, then synchronizes using \texttt{\_\_syncthreads()} before computing the output \texttt{c[tid]}. This synchronization enforces a barrier such that:
\vspace{-0.08in}
\[
\forall i,j \quad \texttt{s\_a[i], s\_b[i] loaded} \rightarrow \texttt{c[j] computed}
\]

In the transformed CPU version, the first loop loads all shared memory values, and the second loop performs the computation. This enforces the same causal dependency: no computation of \texttt{c[id[tid]]} begins until all values in \texttt{s\_a} and \texttt{s\_b} are available.

While the GPU code allows concurrent execution within each phase, the CPU code introduces explicit ordering. This stricter ordering is a valid refinement of the GPU semantics. Therefore, the transformation preserves correctness.

%% file: sources/06_optimization.tex
\section{Optimizations}
\label{sec:optimizations}

Fuzzing requires executing programs on a large input set, making high throughput essential for detecting bugs within a reasonable time. To this end, we propose two compiler–runtime co-design optimizations (\textit{PREX} and \textit{AXIPrune}) tailored specifically for \mname. These optimizations are based on a key insight that unlike existing GPU-to-CPU migration, \mname does not require output equivalence, it only requires memory-bugs present in original program to be accurately preserved in the translated programs.

\begin{figure}[h]
    \centering
    \includegraphics[width=1.0\linewidth]{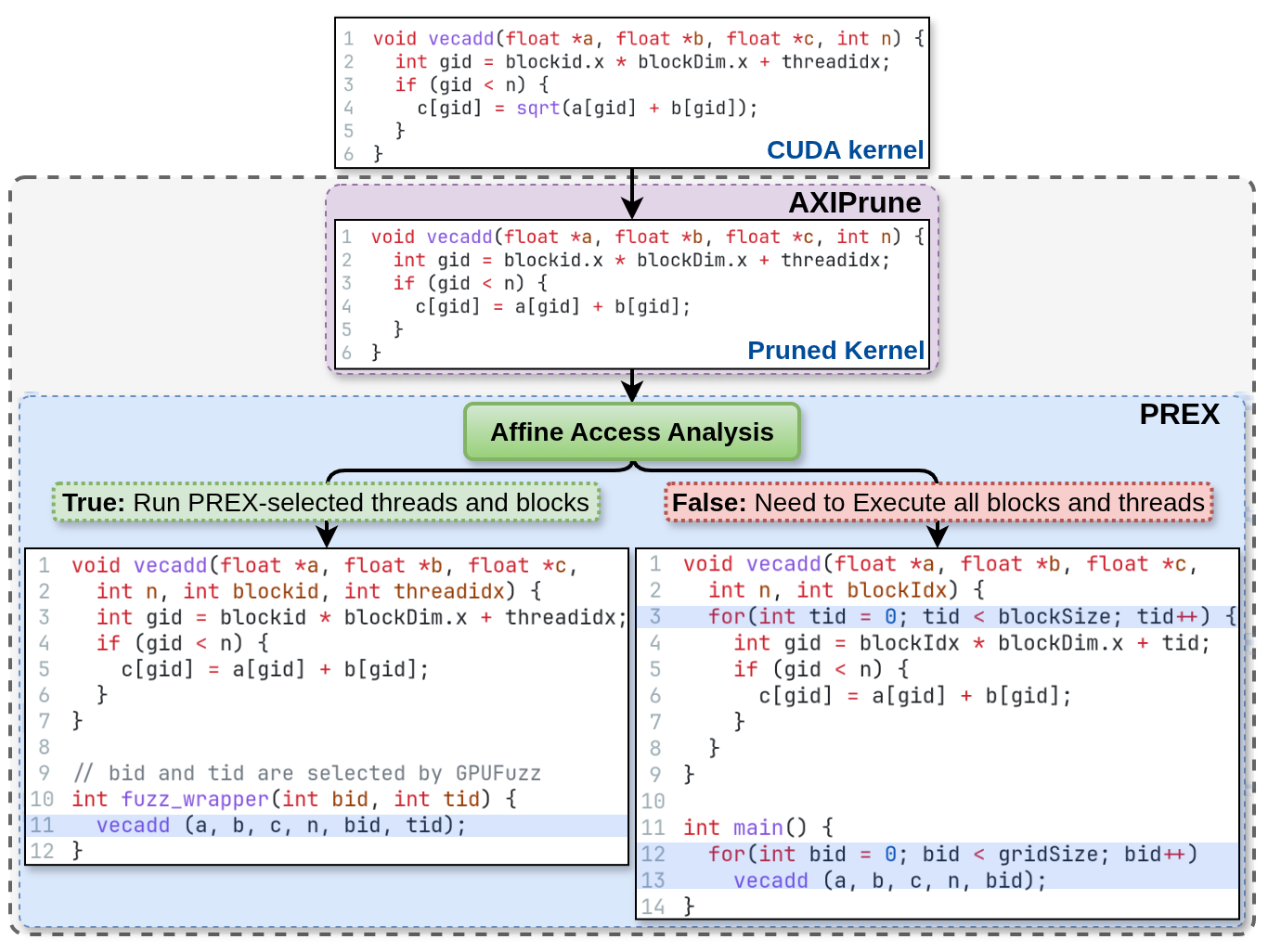}
        \caption{\mname Optimizations}
    \label{fig:GPUFuzz_optimizations}
\end{figure}

\subsection{Partial Representative Execution}
\label{sec:PREX}

The GPU programming model follows Single Program Multiple Data (SPMD) paradigm, where threads usually follow similar control flow. Therefore, in most cases, a GPU memory bug can be detected by executing only a subset of GPU threads. Based on this insight, we propose Partial Representative Execution (\textit{PREX}), a compiler–runtime co-optimization that reduces the number of GPU blocks and threads executed on CPUs in order to improve \mname throughput.

\begin{figure}[htbp]
    \centering
    \includegraphics[width=0.8\linewidth]{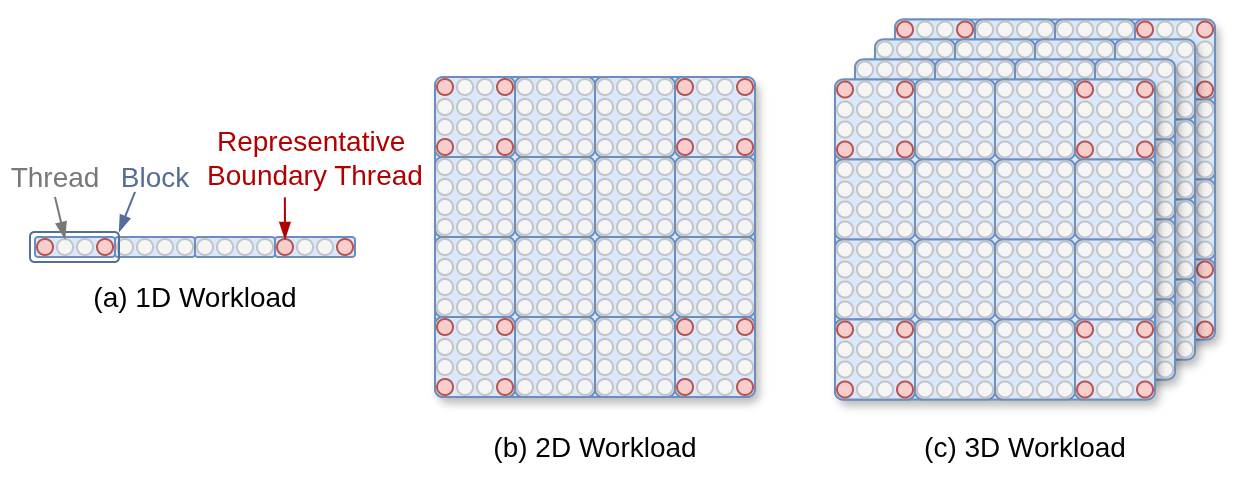}
    \caption{Boundary threads in affine-access kernels}
    \label{fig:prex_boundary_threads_visualization}
\end{figure}

The compiler performs a static analysis (\textit{Affine Access Analysis}) to examine the memory access pattern of the GPU program. If the program is classified as \textit{affine-access} (Section~\ref{sec:affine_access}), PREX generates CPU code that exposes the GPU block and thread indices as function arguments (Figure~\ref{fig:GPUFuzz_optimizations}, bottom left). Hence, \mname executes only a subset of representative blocks and threads at runtime as shown in Fig.~\ref{fig:prex_boundary_threads_visualization}

The PREX runtime is integrated into the \mname runtime. During each fuzzing iteration, it dynamically selects which blocks to execute, prioritizing blocks that are most likely to trigger memory bugs. If no bugs are detected and fuzzing coverage reaches 100\%, PREX runtime terminates the fuzzing process and skips all remaining blocks. By executing only a representative subset of blocks, PREX significantly improves throughput. For non-\textit{affine-access} programs, PREX falls back to executing all GPU blocks and threads to ensure that every thread is examined for potential memory bugs.

\subsubsection{Affine Access Analysis}
\label{sec:affine_access}
GPU programs, due to the SPMD programming model, use block and thread indices to compute memory accesses. In real-world GPU kernels, these accesses are derived from \emph{affine} functions of the indices; we refer to  such CUDA programs \emph{affine-access} programs.

\noindent
\textbf{Definition of Affine Access}: A GPU kernel containing \( M \) memory access instructions is defined as \textit{affine-access} if the memory access indices—represented as a column vector and denoted by \( \overrightarrow{indices} \)—can be expressed as an affine transformation of the block and thread indices. Specifically, the transformation matrix (\( \mathbf{T} \)) and the translation vector (\( \overrightarrow{b} \)) must consist of values that are invariant across all threads (e.g., constants, kernel arguments, or global variables). This relationship is mathematically expressed as:
\[
\overrightarrow{indices} = \mathbf{T} 
\begin{bmatrix}
\text{threadId} \\
\text{blockId}
\end{bmatrix}
+ \overrightarrow{b}.
\]

This formulation extends to multi-dimensional grids and blocks, but for clarity we restrict our discussion to the one-dimensional case.

The kernel in Listing~\ref{lst:pact_orig_gpu_prog} has seven memory accesses using {\tt threadIdx.x} and {\tt id} as indices. 
After omitting duplicates, these indices can be expressed as:
\[
\overrightarrow{indices} =
\begin{bmatrix}
\text{threadIdx.x} \\
\text{blockIdx.x*blockDim.x+threadIdx.x}
\end{bmatrix}.
\]

This can be decomposed into the affine transformation:
\[
\overrightarrow{indices} =
\begin{bmatrix}
1 & 0 \\
1 & \text{blockDim.x}
\end{bmatrix} 
\begin{bmatrix}
\text{threadIdx.x} \\
\text{blockIdx.x}
\end{bmatrix} 
+
\begin{bmatrix}
0 \\
0
\end{bmatrix}.
\]

Since the indices are affine functions of block and thread indices with constant coefficients across all threads, 
this kernel is classified as an \texttt{affine-access} kernel.

We observe that a significant number of real-world GPU programs are affine-access programs. This phenomenon can be attributed to two main reasons. First, since GPUs do not enforce a specific execution order, developers must ensure that multiple threads do not access the same memory address simultaneously. Second, in affine-access programs where the transformation matrix has a value of 1 in the first column, threads with consecutive thread indices access sequential memory addresses. This pattern facilitates memory coalescing to reduce memory access overhead.

For affine-access GPU kernels, executing only a subset of GPU threads is sufficient to detect memory bugs. We present the following theorem and provide a proof to support it.

\noindent
\textbf{Theorem:} \textit{
Consider an affine-access GPU kernel with \( B \) blocks and \( T \) threads per block, and assuming none of the memory accesses are enclosed within conditional statements. Let \( t_i^j \) denote the thread with thread index \( i \) and block index \( j \). Define \texttt{bug\_threads} as the set of threads that trigger memory bugs. If \( \texttt{bug\_threads} \neq \varnothing \), then:
\[
\texttt{bug\_threads} \cap \{t_0^0, t_0^{B-1}, t_{T-1}^0, t_{T-1}^{B-1}\} \neq \varnothing.
\]
}

\noindent
\textbf{Proof:} For memory bugs that are not dependent on thread or block indices, such as double-free (DF), all threads execute the same instructions and produce identical memory behaviors. Therefore, such bugs will also be triggered by threads in the set \( \{t_0^0, t_0^{B-1}, t_{T-1}^0, t_{T-1}^{B-1}\} \).

For out-of-bounds bugs, since \texttt{bug\_threads} is non-empty, there exists at least one thread \( t_i^j \) that triggers the bug. In affine-access kernels, the memory access index can be expressed as \( M i + N j + C \), where \( M \), \( N \), and \( C \) are constants shared across all threads. Without loss of generality, assume \( M, N \geq 0 \), and that an out-of-bound access occurs when \( M i + N j + C < 0 \). For thread \( t_0^0 \), the memory access index evaluates to \( C \). Since \( C \leq M i + N j + C < 0 \), it follows that \( C < 0 \), meaning thread \( t_0^0 \) also triggers the out-of-bounds bug. Furthermore, under the assumption that all memory access instructions are not guarded by conditional statements, all threads execute the same memory operations. Therefore, \( t_0^0 \in \texttt{bug\_threads} \).

The proof for other cases (e.g., \( M \) or \( N \) being negative) follows in a similar manner. By symmetry, the same reasoning applies to \( t_0^{B-1} \), \( t_{T-1}^0 \), and \( t_{T-1}^{B-1} \). Therefore, we conclude that if there exists a thread that triggers a memory bug, then at least one thread in the set \( \{t_0^0, t_0^{B-1}, t_{T-1}^0, t_{T-1}^{B-1}\} \) also trigger the same memory bug. \hfill

From the theorem, it follows that for any affine-access GPU kernel in which memory accesses are not enclosed within conditional statements, executing only the first and last threads of the first and last blocks is sufficient to detect memory bugs. If none of these threads trigger any bugs, it guarantees that all other threads are also free of such bugs. 

However, in real GPU programs, we observe that a large number of memory accesses are enclosed within conditional statements. To further improve throughput for such programs, we introduce additional runtime support in Section~\ref{sec:kernel_fuzzing}.

\subsubsection{Runtime Partial Execution}
\label{sec:kernel_fuzzing}

For an affine-access GPU kernel whose memory accesses are enclosed within conditional statements, executing only the first and last threads in the first and last blocks may not be sufficient to detect memory bugs. For example, for GPU kernel in Listing~\ref{lst:variant_statement}, an out-of-bound memory access occurs when executing the thread with \( \texttt{blockIdx.x} = \lceil N / \texttt{blockSize} \rceil - 1 \) and \( \texttt{threadIdx.x} = \texttt{blockSize} - 2 \). However, when executing the last thread (i.e., \( \texttt{threadIdx.x} = \texttt{blockSize} - 1 \)), the conditional statement (line 3) prevents the memory access, and thus the bug is not triggered.

\begin{lstlisting}[language=C++, caption=GPU Kernels with thread-variant conditions, label={lst:variant_statement}]
__global__ void vecAdd(int *a, int *b, int *c, int N) {
  int id = blockIdx.x * blockSize + threadIdx.x;
  if(id<N) // conditional statement
    c[id+1] = a[id+1] + b[id+1];
}
int main() {
  vecAdd<<<ceil(N/blockSize), blockSize>>>(d_a, d_b, d_c, N);
}
\end{lstlisting}

\begin{algorithm}[htbp]
\small
\caption{PREX Runtime Algorithm}
\label{alg:GPUFuzzRuntime}
\begin{algorithmic}[1]

\Procedure{PREXRuntime}{$\mathcal{P}_{gpu}, n_{blocks}$}
    \State $\textit{head} \gets 0$
    \State $\textit{tail} \gets n_{blocks} - 1$
    \State $\textit{hasBug} \gets \text{False}$
    \While{$\textit{head} \leq \textit{tail}$}
        \If{$\textsc{Coverage}(\mathcal{P}_{gpu}) = 100\% \lor \textit{hasBug}$}
            \State \textbf{break}
        \EndIf
        \State $\textit{hasBug} \gets \textit{hasBug} \lor \textsc{DetectBlock}(\mathcal{P}_{gpu}, \textit{head})$
        \State $\textit{hasBug} \gets \textit{hasBug} \lor \textsc{DetectBlock}(\mathcal{P}_{gpu}, \textit{tail})$
        \State $\textit{head} \gets \textit{head} + 1$
        \State $\textit{tail} \gets \textit{tail} - 1$
    \EndWhile
    \State \Return $\textit{hasBug}$
\EndProcedure

\end{algorithmic}
\end{algorithm}

To support these GPU programs, we propose the PREX runtime (Alg.~\ref{alg:GPUFuzzRuntime}). PREX runtime leverages the fuzzing profiling toolkit (line 6) to track coverage data. It continues to select blocks that have not yet been executed and are most likely to trigger memory bugs (lines 9–10), repeating this process until all memory accesses are covered.

Specifically, in the first iteration, the PREX runtime executes the GPU program using the first and last GPU blocks, with each block executing all of its GPU threads. After execution, if all memory access instructions have been executed, the PREX runtime terminates the fuzzing process, as executing the remaining GPU blocks would not trigger additional memory bugs (as discussed in Sec.~\ref{sec:affine_access}). Otherwise, in the next iteration, it proceeds by launching the second first/last GPU blocks—those most likely to trigger memory bugs, as they contain boundary indices that have not yet been executed.

The PREX runtime is integrated into the \mname mutator (Figure~\ref{fig:gpufuzz_workflow}), which includes a coverage monitor and is responsible for selecting the GPU blocks to execute in each fuzzing iteration.

\subsection{Access-Index Preserving Pruning}
\label{sec:aggressive_pruning}

Access-Index Preserving Pruning (AXIPrune) is a compiler optimization that can be applied alongside PREX to further improve \mname throughput. The key insight is that, while GPU programs contain a large number of computational instructions that introduce significant overhead, many of these instructions are unrelated to memory access behavior. For example, GPU programs typically include numerous floating-point calculations, but these values generally do not influence the indices used for memory accesses.

\begin{figure}[h]
    \centering
    \includegraphics[width=1.0\linewidth]{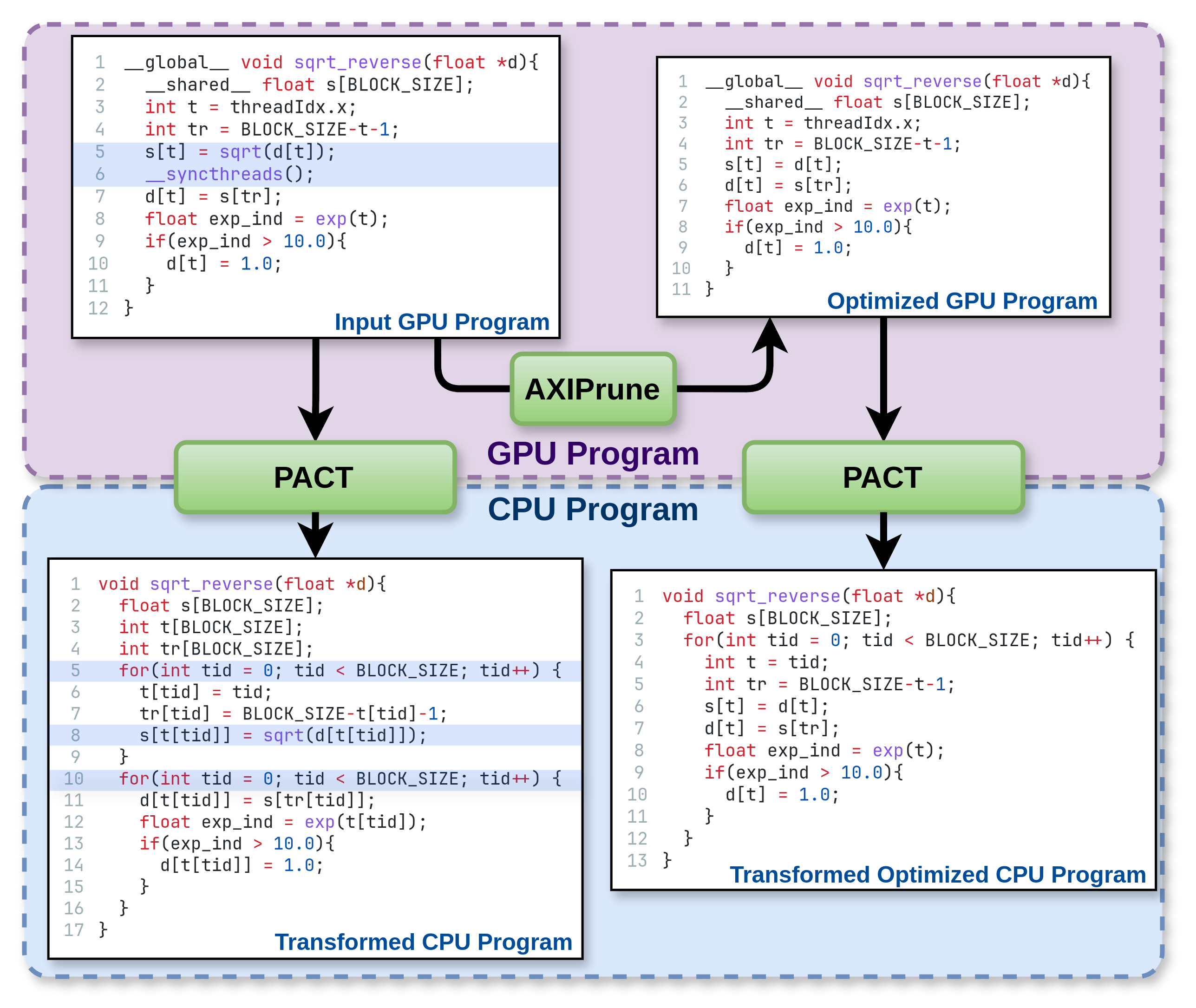}
    \caption{Access-Index Preserving Pruning (AXIPrune)}
    \label{fig:AXIPrune}
\end{figure}

In addition to computation instructions, GPU programs often include barrier instructions (e.g., \texttt{\_\_syncthreads()}) to synchronize access to shared memory. During GPU-to-CPU migration, these barrier instructions are transformed into separate loops, as shown in Listing~\ref{lst:pact_transformed_gpu_prog}. Compared to a single-loop structure, the additional loops introduced by barriers significantly increase execution overhead.

Both computation instructions and barrier instructions contribute significantly to execution overhead. However, in most cases, these instructions do not affect memory access behavior. Their primary role is to ensure numerical correctness rather than to influence memory access indices.

Thus, we propose a compiler optimization, {\tt AXIPrune}, which prunes instructions that are not related to memory access indices. AXIPrune is a pruning solution \textbf{tailored} for \mname and is not intended for general GPU-to-CPU migration. Specifically, AXIPrune may transform a GPU program into one with different semantics that produces different numerical results. However, all memory accesses are preserved exactly as in the original GPU program. The detailed algorithm for AXIPrune is provided in Alg.~\ref{alg:AXIPrune}.

\begin{algorithm}[htb]
\small
\caption{AXIPrune Algorithm}
\label{alg:AXIPrune}
\begin{algorithmic}[1]

\Procedure{AXIPrune}{$\mathcal{P}_{gpu}$}
    \State $\mathcal{P}' \gets \textsc{BarrierElimination}(\mathcal{P}_{gpu})$
    \State $\mathcal{P}' \gets \textsc{MathElimination}(\mathcal{P}')$
    \State \Return $\mathcal{P}'$
\EndProcedure

\Procedure{BarrierElimination}{$\mathcal{P}_{gpu}$}
    \State $\mathcal{S} \gets \text{SharedVars}(\mathcal{P}_{gpu})$
    \State $\textit{memIndexRelated} \gets \text{False}$
    \ForAll{$v \in \mathcal{S}$}
        \If{$\textsc{UsedInBranch}(v) \lor \textsc{UsedAsIndex}(v)$}
            \State $\textit{memIndexRelated} \gets \text{True}$
        \EndIf
    \EndFor
    \If{$\neg \textit{memIndexRelated}$}
        \State $\mathcal{P}_{gpu} \gets \text{RemoveBarrier}(\mathcal{P}_{gpu})$
    \EndIf
    \State \Return $\mathcal{P}_{gpu}$
\EndProcedure

\Procedure{MathElimination}{$\mathcal{P}_{gpu}$}
    \State $\mathcal{M} \gets \text{MathOps}(\mathcal{P}_{gpu})$
    \ForAll{$op \in \mathcal{M}$}
        \If{$\neg(\textsc{UsedInBranch}(op)\lor\textsc{UsedAsIndex}(op))$}
            \State $\mathcal{P}_{gpu} \gets \text{ReplaceWith}(\mathcal{P}_{gpu}, op.output, op.input)$
        \EndIf
    \EndFor
    \State \Return $\mathcal{P}_{gpu}$
\EndProcedure

\end{algorithmic}
\end{algorithm}

AXIPrune is applied to GPU programs and generates a new version with reduced instructions. For example (Figure~\ref{fig:AXIPrune}), the original input GPU program includes a barrier to synchronize access to shared memory. As a result, the transformed CPU program contains two for-loops (lines 5 and 10) that wrap the instructions before and after the barrier. Since the local variables \( t \) and \( tr \) are referenced in both loops, they are promoted to arrays (lines 3–4) with a length equal to the block size.

AXIPrune contains two functions. The first function (Alg.~\ref{alg:AXIPrune}, line 6) eliminates barriers that do not involve memory access indices. In GPU programs, barrier instructions are used to enforce memory consistency for shared memory variables. AXIPrune checks whether these shared memory variables—or the values that depend on them—are used in branch instructions or memory access instructions as index values (Alg.~\ref{alg:AXIPrune}, line 10). If not, then these shared memory variables do not affect memory access behavior, and the barrier instructions can be safely removed.

In the input GPU program in Figure~\ref{fig:AXIPrune}, the memory indices (\( t \) and \( tr \)) are not related to the shared memory values (\( s \)). Thus, AXIPrune removes the barrier (line 6 of the input GPU program). By eliminating the barrier, the transformed optimized CPU program contains only a single for-loop, and the scalar variables \( t \) and \( tr \) no longer need to be extended into arrays, as they are in the unoptimized CPU program (Figure~\ref{fig:AXIPrune}, transformed CPU program, lines 3–4).

The second function in AXIPrune (Alg.~\ref{alg:AXIPrune}, line 19) eliminates expensive math functions that are not related to memory access indices. For GPU program in Figure~\ref{fig:AXIPrune}, the \texttt{sqrt} function (line 5) is unrelated to the memory access indices \( t \) and \( tr \). Therefore, AXIPrune removes this expensive math function in the optimized GPU program. 

In contrast, the other math function, \( \texttt{exp} \), produces an output that is used in a branch instruction (Figure~\ref{fig:AXIPrune}, input GPU program, line 9), which in turn influences the memory access instruction in line 10. To avoid altering memory access behavior, AXIPrune preserves all math functions that are related to memory access indices (Alg.~\ref{alg:AXIPrune}, line 22).

%% file: sources/08_gms.tex
\section{GPU Memory Safety Benchmark}
\label{sec:GMSBench}
\begin{figure}[h]
    \centering
    \includegraphics[width=1.0\linewidth]{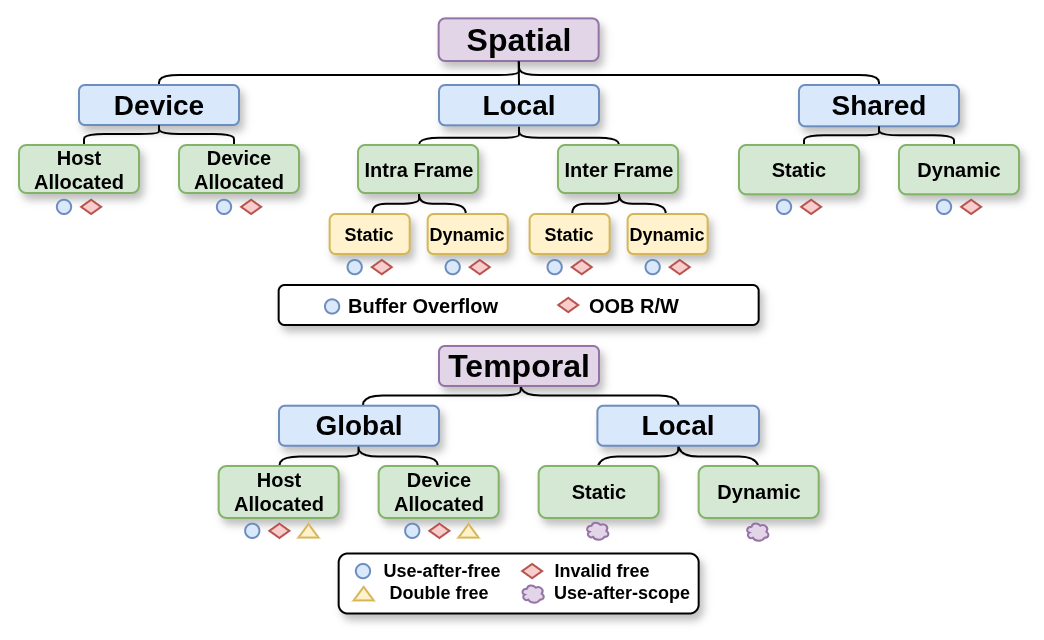}
    \caption{Classification of GPU Memory Safety Bugs}
    \label{fig:GMSBench_classification}
\end{figure}

To evaluate our approach, we developed \textbf{GMSBench}, an CUDA-based opensource GPU Memory Safety Benchmark, inspired by prior work~\cite{ref_cucatch}. \footnote{The CuCatch benchmark is not open source, as it was developed by NVIDIA. In contrast, GMSBench will be made open source with this paper, offering a much larger set of test cases.} It provides a comprehensive framework for evaluating and comparing GPU memory safety techniques. GMSBench test cases are classified by vulnerability class (spatial or temporal), memory region affected (global, local, and shared), type of allocation, and vulnerability type, as shown in Figure~\ref{fig:GMSBench_classification}.

\subsection{Spatial Memory Safety}

The CUDA memory model utilizes global, local, and shared memory spaces. GMSBench includes test cases to detect buffer overflows (BO) and out-of-bounds reads/writes (OOBRW) within each of these memory categories.

Global memory refers to the memory space that all blocks/threads can access. We classify them into global::host-side memory allocations (heap allocation using \texttt{cudaMalloc()}), and global::device-side memory allocations (using \texttt{malloc()} in kernel). BO and OOBRW can occur between two adjacent or non-adjacent allocations and may also extend to other memory spaces such as local and device. 

Local memory refers to thread-local storage. Local memory tests are categorized based on the access domain into intra-buffer accesses (within a stack frame) and inter-buffer accesses (accesses across multiple stack frames). Intra-frame stack allocations can be either static (kernel variables without \texttt{\_\_shared\_\_} attribute) or dynamic (using \texttt{alloca()}). BO and OOBRW may occur between adjacent or non-adjacent allocations and can extend across static and dynamic intra-frame buffers. Finally, local memory accesses (static/dynamic) can also cross into other memory spaces such as global::host-allocated and global::device-allocated.

Shared memory refers to memory shared among threads in a thread block. Shared memory tests are classified based on allocation mechanism into shared::static (allocated using \texttt{\_\_shared\_\_}) and shared::dynamic (allocated during kernel launch). For both types, BO and OOBRW can occur between adjacent and non-adjacent buffers. Additionally, these accesses may also overflow into local (dynamic/static) and global (host/device-allocated) memory regions.

\subsection{Temporal Memory Safety}
Temporal memory safety errors occur when a program accesses memory that has been deallocated or is no longer valid. GMSBench identifies and categorizes these temporal safety cases across different memory regions, specifically focusing on global and local memory. Shared memory is susceptible because it cannot be deallocated within a kernel lifespan. 

Global memory includes host-allocated memory (allocated using \texttt{cudaMalloc()}) and device-allocated memory (allocated dynamically created within kernels using \texttt{malloc()}. Both are prone to UAF errors, where memory is accessed after an object is deallocated. UAF vulnerabilities are classified as immediate and delayed. Immediate UAF arises when an object is accessed immediately after its deallocation, whereas delayed UAF occurs when the object is accessed after a long time, allowing the runtime to potentially reallocate the corresponding memory chunk. Global memory buffers are also susceptible to IF vulnerabilities, which arise when there is a mismatch between the allocator-deallocator calls or the pointer used for deallocation does not point to a valid allocation. Additionally, global memory buffers are also prone to DF vulnerabilities, where the same memory chunk is deallocated more than once, leading to undefined behavior.

Local memory allocations (both static and dynamic) are susceptible to UAS vulnerabilities, where memory is accessed after going out of scope. UAS can be immediate—access right after scope ends—or delayed—access after the memory may be repurposed. Both can cause stack corruption or invalid pointer accesses across function calls.

%% file: sources/07_eval.tex
\section{Evaluation}
\label{sec:eval}

\subsection{Evaluation Methodology}

Our experiments were conducted on a 12th-gen Intel Core i7 with 32 GB RAM and an NVIDIA GeForce RTX 3050 GPU. We used LLVM 14.1 with the ASan runtime and CUDA toolkit 12.1 for compilation. We evaluate \mname and its optimizations through three analyses:

\niparagraph{(1) Error Coverage Analysis:} We analyze the error coverage of using GPU error detection tools and CPU error detection tools on translated binaries using GMSBench, which contains 100 test cases. Error coverage refers to the classes of vulnerabilities that a particular method can detect. In our analysis, we consider the native GPU case as a baseline. For GPU-only solutions, we evaluate GPU execution with NVIDIA Compute Sanitizer~\cite{compSan}, GMOD~\cite{gmod} and GPUShield~\cite{ref_gpushield}. On the other hand, for GPU to CPU transformed programs, we consider Address Sanitizer~\cite{asan_2012} and No-Fat~\cite{ref_nofat}. Compute Sanitizer and Address Sanitizer use tripwires to detect errors, whereas GPUShield and CuCatch track exact base \& bounds. No-Fat uses binning allocators to keep track of base and bound metadata of allocated buffers. 

\niparagraph{(2) Fuzzing Results:} We evaluate \mname’s bug-finding capability by running fuzzing campaigns on a set of CUDA benchmarks and analyzing the resulting crashes. For fuzzing, we used AFL++~\cite{aflpp} (an active fork of the original AFL fuzzer) in LLVM-PCGuard instrumentation mode with recommended settings for performance~\cite{aflpp_perf_tips}. 
 
We fuzzed a set of 14 benchmark programs selected from three benchmark suites:  HeteroMark~\cite{heteromark}, Rodinia~\cite{rodinia} and CUDA SDK~\cite{cuda_sample}. 
All benchmarks were fuzzed for 12 hours with a hang timeout of 10 seconds per execution. For each benchmark, we provided a seed file to the fuzzer without any fine-tuning. All the inputs were mutated solely by AFL++ without any help from the user. AFLTriage~\cite{afltriage} was used to find unique crashes from the crashes reported by AFL++. 

\niparagraph{(3) Performance Analysis:} We evaluate fuzzing performance by measuring average throughput over 100 executions, using executions per second (exec/sec) as the primary metric.

\begin{table*}[h]
\centering
\caption{Detailed Security Coverage Comparison among the other solutions}
\label{tab:detailed_coverage}
\resizebox{0.8\textwidth}{!}{%
\begin{tabular}{|ccccc|c|ccccc|cc|}
\hline
\multicolumn{5}{|c|}{} &
   &
  \multicolumn{5}{c|}{\textbf{GPU Only Solutions}} &
  \multicolumn{2}{c|}{\textbf{\mname}} \\ \cline{7-13} 
\multicolumn{5}{|c|}{\multirow{-2}{*}{\textbf{Test Classification}}} &
  \multirow{-2}{*}{\textbf{No. of Tests}} &
  \multicolumn{1}{c|}{\textbf{Baseline$^a$}} &
  \multicolumn{1}{c|}{\textbf{Comp-San}} &
  \multicolumn{1}{c|}{\textbf{GMOD$^b$}} &
  \multicolumn{1}{c|}{\textbf{GPUShield$^b$}} &
  \textbf{cuCatch$^b$} &
  \multicolumn{1}{c|}{\textbf{ASAN}} &
  \textbf{No-Fat$^b$} \\ \hline
\multicolumn{1}{|c|}{} &
  \multicolumn{1}{c|}{\cellcolor[HTML]{C1F0C8}} &
  \multicolumn{2}{c|}{\cellcolor[HTML]{C1F0C8}} &
  \cellcolor[HTML]{C1F0C8}BO &
  \cellcolor[HTML]{C1F0C8}4 &
  \multicolumn{1}{c|}{\cellcolor[HTML]{C1F0C8}0} &
  \multicolumn{1}{c|}{\cellcolor[HTML]{C1F0C8}4} &
  \multicolumn{1}{c|}{\cellcolor[HTML]{C1F0C8}4} &
  \multicolumn{1}{c|}{\cellcolor[HTML]{C1F0C8}4} &
  \cellcolor[HTML]{C1F0C8}4 &
  \multicolumn{1}{c|}{\cellcolor[HTML]{C1F0C8}4} &
  \cellcolor[HTML]{C1F0C8}4 \\ \cline{5-13} 
\multicolumn{1}{|c|}{} &
  \multicolumn{1}{c|}{\cellcolor[HTML]{C1F0C8}} &
  \multicolumn{2}{c|}{\multirow{-2}{*}{\cellcolor[HTML]{C1F0C8}Alloc. By Host}} &
  \cellcolor[HTML]{C1F0C8}RW &
  \cellcolor[HTML]{C1F0C8}4 &
  \multicolumn{1}{c|}{\cellcolor[HTML]{C1F0C8}0} &
  \multicolumn{1}{c|}{\cellcolor[HTML]{C1F0C8}0} &
  \multicolumn{1}{c|}{\cellcolor[HTML]{C1F0C8}0} &
  \multicolumn{1}{c|}{\cellcolor[HTML]{C1F0C8}4} &
  \cellcolor[HTML]{C1F0C8}4 &
  \multicolumn{1}{c|}{\cellcolor[HTML]{C1F0C8}0} &
  \cellcolor[HTML]{C1F0C8}4 \\ \cline{3-13} 
\multicolumn{1}{|c|}{} &
  \multicolumn{1}{c|}{\cellcolor[HTML]{C1F0C8}} &
  \multicolumn{2}{c|}{\cellcolor[HTML]{C1F0C8}} &
  \cellcolor[HTML]{C1F0C8}BO &
  \cellcolor[HTML]{C1F0C8}4 &
  \multicolumn{1}{c|}{\cellcolor[HTML]{C1F0C8}0} &
  \multicolumn{1}{c|}{\cellcolor[HTML]{C1F0C8}4} &
  \multicolumn{1}{c|}{\cellcolor[HTML]{C1F0C8}4} &
  \multicolumn{1}{c|}{\cellcolor[HTML]{C1F0C8}2} &
  \cellcolor[HTML]{C1F0C8}0 &
  \multicolumn{1}{c|}{\cellcolor[HTML]{C1F0C8}4} &
  \cellcolor[HTML]{C1F0C8}4 \\ \cline{5-13} 
\multicolumn{1}{|c|}{} &
  \multicolumn{1}{c|}{\cellcolor[HTML]{C1F0C8}} &
  \multicolumn{2}{c|}{\multirow{-2}{*}{\cellcolor[HTML]{C1F0C8}Alloc.   By Device}} &
  \cellcolor[HTML]{C1F0C8}RW &
  \cellcolor[HTML]{C1F0C8}4 &
  \multicolumn{1}{c|}{\cellcolor[HTML]{C1F0C8}0} &
  \multicolumn{1}{c|}{\cellcolor[HTML]{C1F0C8}0} &
  \multicolumn{1}{c|}{\cellcolor[HTML]{C1F0C8}0} &
  \multicolumn{1}{c|}{\cellcolor[HTML]{C1F0C8}2} &
  \cellcolor[HTML]{C1F0C8}0 &
  \multicolumn{1}{c|}{\cellcolor[HTML]{C1F0C8}0} &
  \cellcolor[HTML]{C1F0C8}4 \\ \cline{3-13} 
\multicolumn{1}{|c|}{} &
  \multicolumn{1}{c|}{\multirow{-5}{*}{\cellcolor[HTML]{C1F0C8}Global}} &
  \multicolumn{3}{c|}{\cellcolor[HTML]{83E28E}\textbf{Total}} &
  \cellcolor[HTML]{83E28E}\textbf{16} &
  \multicolumn{1}{c|}{\cellcolor[HTML]{83E28E}\textbf{0}} &
  \multicolumn{1}{c|}{\cellcolor[HTML]{83E28E}\textbf{8}} &
  \multicolumn{1}{c|}{\cellcolor[HTML]{83E28E}\textbf{8}} &
  \multicolumn{1}{c|}{\cellcolor[HTML]{83E28E}\textbf{12}} &
  \cellcolor[HTML]{83E28E}\textbf{8} &
  \multicolumn{1}{c|}{\cellcolor[HTML]{83E28E}\textbf{8}} &
  \cellcolor[HTML]{83E28E}\textbf{16} \\ \cline{2-13} 
\multicolumn{1}{|c|}{} &
  \multicolumn{1}{c|}{\cellcolor[HTML]{FBE2D5}} &
  \multicolumn{1}{c|}{\cellcolor[HTML]{FBE2D5}} &
  \multicolumn{1}{c|}{\cellcolor[HTML]{FBE2D5}} &
  \cellcolor[HTML]{FBE2D5}BO &
  \cellcolor[HTML]{FBE2D5}3 &
  \multicolumn{1}{c|}{\cellcolor[HTML]{FBE2D5}0} &
  \multicolumn{1}{c|}{\cellcolor[HTML]{FBE2D5}0} &
  \multicolumn{1}{c|}{\cellcolor[HTML]{FBE2D5}0} &
  \multicolumn{1}{c|}{\cellcolor[HTML]{FBE2D5}0} &
  \cellcolor[HTML]{FBE2D5}0 &
  \multicolumn{1}{c|}{\cellcolor[HTML]{FBE2D5}3} &
  \cellcolor[HTML]{FBE2D5}3 \\ \cline{5-13} 
\multicolumn{1}{|c|}{} &
  \multicolumn{1}{c|}{\cellcolor[HTML]{FBE2D5}} &
  \multicolumn{1}{c|}{\cellcolor[HTML]{FBE2D5}} &
  \multicolumn{1}{c|}{\multirow{-2}{*}{\cellcolor[HTML]{FBE2D5}Static}} &
  \cellcolor[HTML]{FBE2D5}RW &
  \cellcolor[HTML]{FBE2D5}3 &
  \multicolumn{1}{c|}{\cellcolor[HTML]{FBE2D5}0} &
  \multicolumn{1}{c|}{\cellcolor[HTML]{FBE2D5}0} &
  \multicolumn{1}{c|}{\cellcolor[HTML]{FBE2D5}0} &
  \multicolumn{1}{c|}{\cellcolor[HTML]{FBE2D5}0} &
  \cellcolor[HTML]{FBE2D5}0 &
  \multicolumn{1}{c|}{\cellcolor[HTML]{FBE2D5}0} &
  \cellcolor[HTML]{FBE2D5}3 \\ \cline{4-13} 
\multicolumn{1}{|c|}{} &
  \multicolumn{1}{c|}{\cellcolor[HTML]{FBE2D5}} &
  \multicolumn{1}{c|}{\cellcolor[HTML]{FBE2D5}} &
  \multicolumn{1}{c|}{\cellcolor[HTML]{FBE2D5}} &
  \cellcolor[HTML]{FBE2D5}BO &
  \cellcolor[HTML]{FBE2D5}3 &
  \multicolumn{1}{c|}{\cellcolor[HTML]{FBE2D5}0} &
  \multicolumn{1}{c|}{\cellcolor[HTML]{FBE2D5}0} &
  \multicolumn{1}{c|}{\cellcolor[HTML]{FBE2D5}0} &
  \multicolumn{1}{c|}{\cellcolor[HTML]{FBE2D5}0} &
  \cellcolor[HTML]{FBE2D5}3 &
  \multicolumn{1}{c|}{\cellcolor[HTML]{FBE2D5}3} &
  \cellcolor[HTML]{FBE2D5}3 \\ \cline{5-13} 
\multicolumn{1}{|c|}{} &
  \multicolumn{1}{c|}{\cellcolor[HTML]{FBE2D5}} &
  \multicolumn{1}{c|}{\multirow{-4}{*}{\cellcolor[HTML]{FBE2D5}Intra frame}} &
  \multicolumn{1}{c|}{\multirow{-2}{*}{\cellcolor[HTML]{FBE2D5}Dynamic}} &
  \cellcolor[HTML]{FBE2D5}RW &
  \cellcolor[HTML]{FBE2D5}3 &
  \multicolumn{1}{c|}{\cellcolor[HTML]{FBE2D5}0} &
  \multicolumn{1}{c|}{\cellcolor[HTML]{FBE2D5}0} &
  \multicolumn{1}{c|}{\cellcolor[HTML]{FBE2D5}0} &
  \multicolumn{1}{c|}{\cellcolor[HTML]{FBE2D5}0} &
  \cellcolor[HTML]{FBE2D5}3 &
  \multicolumn{1}{c|}{\cellcolor[HTML]{FBE2D5}0} &
  \cellcolor[HTML]{FBE2D5}3 \\ \cline{3-13} 
\multicolumn{1}{|c|}{} &
  \multicolumn{1}{c|}{\cellcolor[HTML]{FBE2D5}} &
  \multicolumn{1}{c|}{\cellcolor[HTML]{FBE2D5}} &
  \multicolumn{1}{c|}{\cellcolor[HTML]{FBE2D5}} &
  \cellcolor[HTML]{FBE2D5}BO &
  \cellcolor[HTML]{FBE2D5}2 &
  \multicolumn{1}{c|}{\cellcolor[HTML]{FBE2D5}0} &
  \multicolumn{1}{c|}{\cellcolor[HTML]{FBE2D5}0} &
  \multicolumn{1}{c|}{\cellcolor[HTML]{FBE2D5}0} &
  \multicolumn{1}{c|}{\cellcolor[HTML]{FBE2D5}0} &
  \cellcolor[HTML]{FBE2D5}0 &
  \multicolumn{1}{c|}{\cellcolor[HTML]{FBE2D5}2} &
  \cellcolor[HTML]{FBE2D5}2 \\ \cline{5-13} 
\multicolumn{1}{|c|}{} &
  \multicolumn{1}{c|}{\cellcolor[HTML]{FBE2D5}} &
  \multicolumn{1}{c|}{\cellcolor[HTML]{FBE2D5}} &
  \multicolumn{1}{c|}{\multirow{-2}{*}{\cellcolor[HTML]{FBE2D5}Static}} &
  \cellcolor[HTML]{FBE2D5}RW &
  \cellcolor[HTML]{FBE2D5}2 &
  \multicolumn{1}{c|}{\cellcolor[HTML]{FBE2D5}0} &
  \multicolumn{1}{c|}{\cellcolor[HTML]{FBE2D5}0} &
  \multicolumn{1}{c|}{\cellcolor[HTML]{FBE2D5}0} &
  \multicolumn{1}{c|}{\cellcolor[HTML]{FBE2D5}0} &
  \cellcolor[HTML]{FBE2D5}0 &
  \multicolumn{1}{c|}{\cellcolor[HTML]{FBE2D5}0} &
  \cellcolor[HTML]{FBE2D5}2 \\ \cline{4-13} 
\multicolumn{1}{|c|}{} &
  \multicolumn{1}{c|}{\cellcolor[HTML]{FBE2D5}} &
  \multicolumn{1}{c|}{\cellcolor[HTML]{FBE2D5}} &
  \multicolumn{1}{c|}{\cellcolor[HTML]{FBE2D5}} &
  \cellcolor[HTML]{FBE2D5}BO &
  \cellcolor[HTML]{FBE2D5}2 &
  \multicolumn{1}{c|}{\cellcolor[HTML]{FBE2D5}0} &
  \multicolumn{1}{c|}{\cellcolor[HTML]{FBE2D5}0} &
  \multicolumn{1}{c|}{\cellcolor[HTML]{FBE2D5}0} &
  \multicolumn{1}{c|}{\cellcolor[HTML]{FBE2D5}0} &
  \cellcolor[HTML]{FBE2D5}2 &
  \multicolumn{1}{c|}{\cellcolor[HTML]{FBE2D5}2} &
  \cellcolor[HTML]{FBE2D5}2 \\ \cline{5-13} 
\multicolumn{1}{|c|}{} &
  \multicolumn{1}{c|}{\cellcolor[HTML]{FBE2D5}} &
  \multicolumn{1}{c|}{\multirow{-4}{*}{\cellcolor[HTML]{FBE2D5}Inter frame}} &
  \multicolumn{1}{c|}{\multirow{-2}{*}{\cellcolor[HTML]{FBE2D5}Dynamic}} &
  \cellcolor[HTML]{FBE2D5}RW &
  \cellcolor[HTML]{FBE2D5}2 &
  \multicolumn{1}{c|}{\cellcolor[HTML]{FBE2D5}0} &
  \multicolumn{1}{c|}{\cellcolor[HTML]{FBE2D5}0} &
  \multicolumn{1}{c|}{\cellcolor[HTML]{FBE2D5}0} &
  \multicolumn{1}{c|}{\cellcolor[HTML]{FBE2D5}0} &
  \cellcolor[HTML]{FBE2D5}2 &
  \multicolumn{1}{c|}{\cellcolor[HTML]{FBE2D5}0} &
  \cellcolor[HTML]{FBE2D5}2 \\ \cline{3-13} 
\multicolumn{1}{|c|}{} &
  \multicolumn{1}{c|}{\cellcolor[HTML]{FBE2D5}} &
  \multicolumn{1}{c|}{\cellcolor[HTML]{FBE2D5}} &
  \multicolumn{1}{c|}{\cellcolor[HTML]{FBE2D5}} &
  \cellcolor[HTML]{FBE2D5}BO &
  \cellcolor[HTML]{FBE2D5}2 &
  \multicolumn{1}{c|}{\cellcolor[HTML]{FBE2D5}0} &
  \multicolumn{1}{c|}{\cellcolor[HTML]{FBE2D5}2} &
  \multicolumn{1}{c|}{\cellcolor[HTML]{FBE2D5}0} &
  \multicolumn{1}{c|}{\cellcolor[HTML]{FBE2D5}2} &
  \cellcolor[HTML]{FBE2D5}2 &
  \multicolumn{1}{c|}{\cellcolor[HTML]{FBE2D5}2} &
  \cellcolor[HTML]{FBE2D5}2 \\ \cline{5-13} 
\multicolumn{1}{|c|}{} &
  \multicolumn{1}{c|}{\cellcolor[HTML]{FBE2D5}} &
  \multicolumn{1}{c|}{\cellcolor[HTML]{FBE2D5}} &
  \multicolumn{1}{c|}{\multirow{-2}{*}{\cellcolor[HTML]{FBE2D5}Static}} &
  \cellcolor[HTML]{FBE2D5}RW &
  \cellcolor[HTML]{FBE2D5}2 &
  \multicolumn{1}{c|}{\cellcolor[HTML]{FBE2D5}0} &
  \multicolumn{1}{c|}{\cellcolor[HTML]{FBE2D5}2} &
  \multicolumn{1}{c|}{\cellcolor[HTML]{FBE2D5}0} &
  \multicolumn{1}{c|}{\cellcolor[HTML]{FBE2D5}2} &
  \cellcolor[HTML]{FBE2D5}2 &
  \multicolumn{1}{c|}{\cellcolor[HTML]{FBE2D5}0} &
  \cellcolor[HTML]{FBE2D5}2 \\ \cline{4-13} 
\multicolumn{1}{|c|}{} &
  \multicolumn{1}{c|}{\cellcolor[HTML]{FBE2D5}} &
  \multicolumn{1}{c|}{\cellcolor[HTML]{FBE2D5}} &
  \multicolumn{1}{c|}{\cellcolor[HTML]{FBE2D5}} &
  \cellcolor[HTML]{FBE2D5}BO &
  \cellcolor[HTML]{FBE2D5}2 &
  \multicolumn{1}{c|}{\cellcolor[HTML]{FBE2D5}0} &
  \multicolumn{1}{c|}{\cellcolor[HTML]{FBE2D5}2} &
  \multicolumn{1}{c|}{\cellcolor[HTML]{FBE2D5}0} &
  \multicolumn{1}{c|}{\cellcolor[HTML]{FBE2D5}2} &
  \cellcolor[HTML]{FBE2D5}2 &
  \multicolumn{1}{c|}{\cellcolor[HTML]{FBE2D5}2} &
  \cellcolor[HTML]{FBE2D5}2 \\ \cline{5-13} 
\multicolumn{1}{|c|}{} &
  \multicolumn{1}{c|}{\cellcolor[HTML]{FBE2D5}} &
  \multicolumn{1}{c|}{\multirow{-4}{*}{\cellcolor[HTML]{FBE2D5}Beyond Local}} &
  \multicolumn{1}{c|}{\multirow{-2}{*}{\cellcolor[HTML]{FBE2D5}Dynamic}} &
  \cellcolor[HTML]{FBE2D5}RW &
  \cellcolor[HTML]{FBE2D5}2 &
  \multicolumn{1}{c|}{\cellcolor[HTML]{FBE2D5}0} &
  \multicolumn{1}{c|}{\cellcolor[HTML]{FBE2D5}2} &
  \multicolumn{1}{c|}{\cellcolor[HTML]{FBE2D5}0} &
  \multicolumn{1}{c|}{\cellcolor[HTML]{FBE2D5}2} &
  \cellcolor[HTML]{FBE2D5}2 &
  \multicolumn{1}{c|}{\cellcolor[HTML]{FBE2D5}0} &
  \cellcolor[HTML]{FBE2D5}2 \\ \cline{3-13} 
\multicolumn{1}{|c|}{} &
  \multicolumn{1}{c|}{\multirow{-13}{*}{\cellcolor[HTML]{FBE2D5}Local}} &
  \multicolumn{3}{c|}{\cellcolor[HTML]{F7C7AC}\textbf{Total}} &
  \cellcolor[HTML]{F7C7AC}\textbf{28} &
  \multicolumn{1}{c|}{\cellcolor[HTML]{F7C7AC}\textbf{0}} &
  \multicolumn{1}{c|}{\cellcolor[HTML]{F7C7AC}\textbf{8}} &
  \multicolumn{1}{c|}{\cellcolor[HTML]{F7C7AC}\textbf{0}} &
  \multicolumn{1}{c|}{\cellcolor[HTML]{F7C7AC}\textbf{8}} &
  \cellcolor[HTML]{F7C7AC}\textbf{18} &
  \multicolumn{1}{c|}{\cellcolor[HTML]{F7C7AC}\textbf{14}} &
  \cellcolor[HTML]{F7C7AC}\textbf{28} \\ \cline{2-13} 
\multicolumn{1}{|c|}{} &
  \multicolumn{1}{c|}{\cellcolor[HTML]{C0E6F5}} &
  \multicolumn{2}{c|}{\cellcolor[HTML]{C0E6F5}} &
  \cellcolor[HTML]{C0E6F5}BO &
  \cellcolor[HTML]{C0E6F5}7 &
  \multicolumn{1}{c|}{\cellcolor[HTML]{C0E6F5}0} &
  \multicolumn{1}{c|}{\cellcolor[HTML]{C0E6F5}4} &
  \multicolumn{1}{c|}{\cellcolor[HTML]{C0E6F5}0} &
  \multicolumn{1}{c|}{\cellcolor[HTML]{C0E6F5}0} &
  \cellcolor[HTML]{C0E6F5}7 &
  \multicolumn{1}{c|}{\cellcolor[HTML]{C0E6F5}7} &
  \cellcolor[HTML]{C0E6F5}7 \\ \cline{5-13} 
\multicolumn{1}{|c|}{} &
  \multicolumn{1}{c|}{\cellcolor[HTML]{C0E6F5}} &
  \multicolumn{2}{c|}{\multirow{-2}{*}{\cellcolor[HTML]{C0E6F5}Static}} &
  \cellcolor[HTML]{C0E6F5}RW &
  \cellcolor[HTML]{C0E6F5}7 &
  \multicolumn{1}{c|}{\cellcolor[HTML]{C0E6F5}0} &
  \multicolumn{1}{c|}{\cellcolor[HTML]{C0E6F5}0} &
  \multicolumn{1}{c|}{\cellcolor[HTML]{C0E6F5}0} &
  \multicolumn{1}{c|}{\cellcolor[HTML]{C0E6F5}0} &
  \cellcolor[HTML]{C0E6F5}7 &
  \multicolumn{1}{c|}{\cellcolor[HTML]{C0E6F5}0} &
  \cellcolor[HTML]{C0E6F5}7 \\ \cline{3-13} 
\multicolumn{1}{|c|}{} &
  \multicolumn{1}{c|}{\cellcolor[HTML]{C0E6F5}} &
  \multicolumn{2}{c|}{\cellcolor[HTML]{C0E6F5}} &
  \cellcolor[HTML]{C0E6F5}BO &
  \cellcolor[HTML]{C0E6F5}7 &
  \multicolumn{1}{c|}{\cellcolor[HTML]{C0E6F5}0} &
  \multicolumn{1}{c|}{\cellcolor[HTML]{C0E6F5}4} &
  \multicolumn{1}{c|}{\cellcolor[HTML]{C0E6F5}0} &
  \multicolumn{1}{c|}{\cellcolor[HTML]{C0E6F5}0} &
  \cellcolor[HTML]{C0E6F5}5 &
  \multicolumn{1}{c|}{\cellcolor[HTML]{C0E6F5}5} &
  \cellcolor[HTML]{C0E6F5}7 \\ \cline{5-13} 
\multicolumn{1}{|c|}{} &
  \multicolumn{1}{c|}{\cellcolor[HTML]{C0E6F5}} &
  \multicolumn{2}{c|}{\multirow{-2}{*}{\cellcolor[HTML]{C0E6F5}Dynamic}} &
  \cellcolor[HTML]{C0E6F5}RW &
  \cellcolor[HTML]{C0E6F5}7 &
  \multicolumn{1}{c|}{\cellcolor[HTML]{C0E6F5}0} &
  \multicolumn{1}{c|}{\cellcolor[HTML]{C0E6F5}0} &
  \multicolumn{1}{c|}{\cellcolor[HTML]{C0E6F5}0} &
  \multicolumn{1}{c|}{\cellcolor[HTML]{C0E6F5}0} &
  \cellcolor[HTML]{C0E6F5}5 &
  \multicolumn{1}{c|}{\cellcolor[HTML]{C0E6F5}0} &
  \cellcolor[HTML]{C0E6F5}0 \\ \cline{3-13} 
\multicolumn{1}{|c|}{} &
  \multicolumn{1}{c|}{\multirow{-5}{*}{\cellcolor[HTML]{C0E6F5}Shared}} &
  \multicolumn{3}{c|}{\cellcolor[HTML]{83CCEB}\textbf{Total}} &
  \cellcolor[HTML]{83CCEB}\textbf{28} &
  \multicolumn{1}{c|}{\cellcolor[HTML]{83CCEB}\textbf{0}} &
  \multicolumn{1}{c|}{\cellcolor[HTML]{83CCEB}\textbf{8}} &
  \multicolumn{1}{c|}{\cellcolor[HTML]{83CCEB}\textbf{0}} &
  \multicolumn{1}{c|}{\cellcolor[HTML]{83CCEB}\textbf{0}} &
  \cellcolor[HTML]{83CCEB}\textbf{24} &
  \multicolumn{1}{c|}{\cellcolor[HTML]{83CCEB}\textbf{12}} &
  \cellcolor[HTML]{83CCEB}\textbf{21} \\ \cline{2-13} 
\multicolumn{1}{|c|}{} &
  \multicolumn{1}{c|}{\cellcolor[HTML]{DAE9F8}} &
  \multicolumn{2}{c|}{\cellcolor[HTML]{DAE9F8}} &
  \cellcolor[HTML]{DAE9F8}BO &
  \cellcolor[HTML]{DAE9F8}2 &
  \multicolumn{1}{c|}{\cellcolor[HTML]{DAE9F8}0} &
  \multicolumn{1}{c|}{\cellcolor[HTML]{DAE9F8}0} &
  \multicolumn{1}{c|}{\cellcolor[HTML]{DAE9F8}0} &
  \multicolumn{1}{c|}{\cellcolor[HTML]{DAE9F8}0} &
  \cellcolor[HTML]{DAE9F8}0 &
  \multicolumn{1}{c|}{\cellcolor[HTML]{DAE9F8}2} &
  \cellcolor[HTML]{DAE9F8}2 \\ \cline{5-13} 
\multicolumn{1}{|c|}{} &
  \multicolumn{1}{c|}{\cellcolor[HTML]{DAE9F8}} &
  \multicolumn{2}{c|}{\multirow{-2}{*}{\cellcolor[HTML]{DAE9F8}Global}} &
  \cellcolor[HTML]{DAE9F8}RW &
  \cellcolor[HTML]{DAE9F8}2 &
  \multicolumn{1}{c|}{\cellcolor[HTML]{DAE9F8}0} &
  \multicolumn{1}{c|}{\cellcolor[HTML]{DAE9F8}0} &
  \multicolumn{1}{c|}{\cellcolor[HTML]{DAE9F8}0} &
  \multicolumn{1}{c|}{\cellcolor[HTML]{DAE9F8}0} &
  \cellcolor[HTML]{DAE9F8}0 &
  \multicolumn{1}{c|}{\cellcolor[HTML]{DAE9F8}0} &
  \cellcolor[HTML]{DAE9F8}2 \\ \cline{3-13} 
\multicolumn{1}{|c|}{} &
  \multicolumn{1}{c|}{\cellcolor[HTML]{DAE9F8}} &
  \multicolumn{2}{c|}{\cellcolor[HTML]{DAE9F8}} &
  \cellcolor[HTML]{DAE9F8}BO &
  \cellcolor[HTML]{DAE9F8}2 &
  \multicolumn{1}{c|}{\cellcolor[HTML]{DAE9F8}0} &
  \multicolumn{1}{c|}{\cellcolor[HTML]{DAE9F8}0} &
  \multicolumn{1}{c|}{\cellcolor[HTML]{DAE9F8}0} &
  \multicolumn{1}{c|}{\cellcolor[HTML]{DAE9F8}0} &
  \cellcolor[HTML]{DAE9F8}0 &
  \multicolumn{1}{c|}{\cellcolor[HTML]{DAE9F8}2} &
  \cellcolor[HTML]{DAE9F8}2 \\ \cline{5-13} 
\multicolumn{1}{|c|}{} &
  \multicolumn{1}{c|}{\cellcolor[HTML]{DAE9F8}} &
  \multicolumn{2}{c|}{\multirow{-2}{*}{\cellcolor[HTML]{DAE9F8}Local}} &
  \cellcolor[HTML]{DAE9F8}RW &
  \cellcolor[HTML]{DAE9F8}2 &
  \multicolumn{1}{c|}{\cellcolor[HTML]{DAE9F8}0} &
  \multicolumn{1}{c|}{\cellcolor[HTML]{DAE9F8}0} &
  \multicolumn{1}{c|}{\cellcolor[HTML]{DAE9F8}0} &
  \multicolumn{1}{c|}{\cellcolor[HTML]{DAE9F8}0} &
  \cellcolor[HTML]{DAE9F8}0 &
  \multicolumn{1}{c|}{\cellcolor[HTML]{DAE9F8}0} &
  \cellcolor[HTML]{DAE9F8}2 \\ \cline{3-13} 
\multicolumn{1}{|c|}{} &
  \multicolumn{1}{c|}{\cellcolor[HTML]{DAE9F8}} &
  \multicolumn{2}{c|}{\cellcolor[HTML]{DAE9F8}} &
  \cellcolor[HTML]{DAE9F8}BO &
  \cellcolor[HTML]{DAE9F8}2 &
  \multicolumn{1}{c|}{\cellcolor[HTML]{DAE9F8}0} &
  \multicolumn{1}{c|}{\cellcolor[HTML]{DAE9F8}0} &
  \multicolumn{1}{c|}{\cellcolor[HTML]{DAE9F8}0} &
  \multicolumn{1}{c|}{\cellcolor[HTML]{DAE9F8}0} &
  \cellcolor[HTML]{DAE9F8}0 &
  \multicolumn{1}{c|}{\cellcolor[HTML]{DAE9F8}2} &
  \cellcolor[HTML]{DAE9F8}2 \\ \cline{5-13} 
\multicolumn{1}{|c|}{} &
  \multicolumn{1}{c|}{\cellcolor[HTML]{DAE9F8}} &
  \multicolumn{2}{c|}{\multirow{-2}{*}{\cellcolor[HTML]{DAE9F8}Shared}} &
  \cellcolor[HTML]{DAE9F8}RW &
  \cellcolor[HTML]{DAE9F8}2 &
  \multicolumn{1}{c|}{\cellcolor[HTML]{DAE9F8}0} &
  \multicolumn{1}{c|}{\cellcolor[HTML]{DAE9F8}0} &
  \multicolumn{1}{c|}{\cellcolor[HTML]{DAE9F8}0} &
  \multicolumn{1}{c|}{\cellcolor[HTML]{DAE9F8}0} &
  \cellcolor[HTML]{DAE9F8}0 &
  \multicolumn{1}{c|}{\cellcolor[HTML]{DAE9F8}0} &
  \cellcolor[HTML]{DAE9F8}2 \\ \cline{3-13} 
\multicolumn{1}{|c|}{\multirow{-30}{*}{\textbf{Spatial}}} &
  \multicolumn{1}{c|}{\multirow{-7}{*}{\cellcolor[HTML]{DAE9F8}Intra Allocation}} &
  \multicolumn{3}{c|}{\cellcolor[HTML]{A6C9EC}\textbf{Total}} &
  \cellcolor[HTML]{A6C9EC}\textbf{12} &
  \multicolumn{1}{c|}{\cellcolor[HTML]{A6C9EC}\textbf{0}} &
  \multicolumn{1}{c|}{\cellcolor[HTML]{A6C9EC}\textbf{0}} &
  \multicolumn{1}{c|}{\cellcolor[HTML]{A6C9EC}\textbf{0}} &
  \multicolumn{1}{c|}{\cellcolor[HTML]{A6C9EC}\textbf{0}} &
  \cellcolor[HTML]{A6C9EC}\textbf{0} &
  \multicolumn{1}{c|}{\cellcolor[HTML]{A6C9EC}\textbf{6}} &
  \cellcolor[HTML]{A6C9EC}\textbf{12} \\ \hline
\multicolumn{1}{|c|}{} &
  \multicolumn{1}{c|}{\cellcolor[HTML]{C1F0C8}} &
  \multicolumn{2}{c|}{\cellcolor[HTML]{C1F0C8}} &
  \cellcolor[HTML]{C1F0C8}UAF &
  \cellcolor[HTML]{C1F0C8}2 &
  \multicolumn{1}{c|}{\cellcolor[HTML]{C1F0C8}0} &
  \multicolumn{1}{c|}{\cellcolor[HTML]{C1F0C8}1} &
  \multicolumn{1}{c|}{\cellcolor[HTML]{C1F0C8}0} &
  \multicolumn{1}{c|}{\cellcolor[HTML]{C1F0C8}0} &
  \cellcolor[HTML]{C1F0C8}1 &
  \multicolumn{1}{c|}{\cellcolor[HTML]{C1F0C8}1} &
  \cellcolor[HTML]{C1F0C8}1 \\ \cline{5-13} 
\multicolumn{1}{|c|}{} &
  \multicolumn{1}{c|}{\cellcolor[HTML]{C1F0C8}} &
  \multicolumn{2}{c|}{\cellcolor[HTML]{C1F0C8}} &
  \cellcolor[HTML]{C1F0C8}IF &
  \cellcolor[HTML]{C1F0C8}3 &
  \multicolumn{1}{c|}{\cellcolor[HTML]{C1F0C8}3} &
  \multicolumn{1}{c|}{\cellcolor[HTML]{C1F0C8}3} &
  \multicolumn{1}{c|}{\cellcolor[HTML]{C1F0C8}3} &
  \multicolumn{1}{c|}{\cellcolor[HTML]{C1F0C8}3} &
  \cellcolor[HTML]{C1F0C8}3 &
  \multicolumn{1}{c|}{\cellcolor[HTML]{C1F0C8}1} &
  \cellcolor[HTML]{C1F0C8}3 \\ \cline{5-13} 
\multicolumn{1}{|c|}{} &
  \multicolumn{1}{c|}{\cellcolor[HTML]{C1F0C8}} &
  \multicolumn{2}{c|}{\multirow{-3}{*}{\cellcolor[HTML]{C1F0C8}Alloc. By Host}} &
  \cellcolor[HTML]{C1F0C8}DF &
  \cellcolor[HTML]{C1F0C8}1 &
  \multicolumn{1}{c|}{\cellcolor[HTML]{C1F0C8}1} &
  \multicolumn{1}{c|}{\cellcolor[HTML]{C1F0C8}1} &
  \multicolumn{1}{c|}{\cellcolor[HTML]{C1F0C8}1} &
  \multicolumn{1}{c|}{\cellcolor[HTML]{C1F0C8}1} &
  \cellcolor[HTML]{C1F0C8}1 &
  \multicolumn{1}{c|}{\cellcolor[HTML]{C1F0C8}1} &
  \cellcolor[HTML]{C1F0C8}1 \\ \cline{3-13} 
\multicolumn{1}{|c|}{} &
  \multicolumn{1}{c|}{\cellcolor[HTML]{C1F0C8}} &
  \multicolumn{2}{c|}{\cellcolor[HTML]{C1F0C8}} &
  \cellcolor[HTML]{C1F0C8}UAF &
  \cellcolor[HTML]{C1F0C8}2 &
  \multicolumn{1}{c|}{\cellcolor[HTML]{C1F0C8}0} &
  \multicolumn{1}{c|}{\cellcolor[HTML]{C1F0C8}1} &
  \multicolumn{1}{c|}{\cellcolor[HTML]{C1F0C8}0} &
  \multicolumn{1}{c|}{\cellcolor[HTML]{C1F0C8}0} &
  \cellcolor[HTML]{C1F0C8}1 &
  \multicolumn{1}{c|}{\cellcolor[HTML]{C1F0C8}1} &
  \cellcolor[HTML]{C1F0C8}1 \\ \cline{5-13} 
\multicolumn{1}{|c|}{} &
  \multicolumn{1}{c|}{\cellcolor[HTML]{C1F0C8}} &
  \multicolumn{2}{c|}{\cellcolor[HTML]{C1F0C8}} &
  \cellcolor[HTML]{C1F0C8}IF &
  \cellcolor[HTML]{C1F0C8}3 &
  \multicolumn{1}{c|}{\cellcolor[HTML]{C1F0C8}3} &
  \multicolumn{1}{c|}{\cellcolor[HTML]{C1F0C8}3} &
  \multicolumn{1}{c|}{\cellcolor[HTML]{C1F0C8}3} &
  \multicolumn{1}{c|}{\cellcolor[HTML]{C1F0C8}3} &
  \cellcolor[HTML]{C1F0C8}3 &
  \multicolumn{1}{c|}{\cellcolor[HTML]{C1F0C8}1} &
  \cellcolor[HTML]{C1F0C8}3 \\ \cline{5-13} 
\multicolumn{1}{|c|}{} &
  \multicolumn{1}{c|}{\cellcolor[HTML]{C1F0C8}} &
  \multicolumn{2}{c|}{\multirow{-3}{*}{\cellcolor[HTML]{C1F0C8}Alloc.   By Device}} &
  \cellcolor[HTML]{C1F0C8}DF &
  \cellcolor[HTML]{C1F0C8}1 &
  \multicolumn{1}{c|}{\cellcolor[HTML]{C1F0C8}1} &
  \multicolumn{1}{c|}{\cellcolor[HTML]{C1F0C8}1} &
  \multicolumn{1}{c|}{\cellcolor[HTML]{C1F0C8}1} &
  \multicolumn{1}{c|}{\cellcolor[HTML]{C1F0C8}1} &
  \cellcolor[HTML]{C1F0C8}1 &
  \multicolumn{1}{c|}{\cellcolor[HTML]{C1F0C8}1} &
  \cellcolor[HTML]{C1F0C8}1 \\ \cline{3-13} 
\multicolumn{1}{|c|}{} &
  \multicolumn{1}{c|}{\multirow{-7}{*}{\cellcolor[HTML]{C1F0C8}Global}} &
  \multicolumn{3}{c|}{\cellcolor[HTML]{83E28E}\textbf{Total}} &
  \cellcolor[HTML]{83E28E}\textbf{12} &
  \multicolumn{1}{c|}{\cellcolor[HTML]{83E28E}\textbf{8}} &
  \multicolumn{1}{c|}{\cellcolor[HTML]{83E28E}\textbf{10}} &
  \multicolumn{1}{c|}{\cellcolor[HTML]{83E28E}\textbf{8}} &
  \multicolumn{1}{c|}{\cellcolor[HTML]{83E28E}\textbf{8}} &
  \cellcolor[HTML]{83E28E}\textbf{10} &
  \multicolumn{1}{c|}{\cellcolor[HTML]{83E28E}\textbf{6}} &
  \cellcolor[HTML]{83E28E}\textbf{10} \\ \cline{2-13} 
\multicolumn{1}{|c|}{} &
  \multicolumn{1}{c|}{\cellcolor[HTML]{FBE2D5}} &
  \multicolumn{2}{c|}{\cellcolor[HTML]{FBE2D5}Static} &
  \cellcolor[HTML]{FBE2D5}UAS &
  \cellcolor[HTML]{FBE2D5}2 &
  \multicolumn{1}{c|}{\cellcolor[HTML]{FBE2D5}0} &
  \multicolumn{1}{c|}{\cellcolor[HTML]{FBE2D5}1} &
  \multicolumn{1}{c|}{\cellcolor[HTML]{FBE2D5}0} &
  \multicolumn{1}{c|}{\cellcolor[HTML]{FBE2D5}0} &
  \cellcolor[HTML]{FBE2D5}2 &
  \multicolumn{1}{c|}{\cellcolor[HTML]{FBE2D5}1} &
  \cellcolor[HTML]{FBE2D5}2 \\ \cline{3-13} 
\multicolumn{1}{|c|}{} &
  \multicolumn{1}{c|}{\cellcolor[HTML]{FBE2D5}} &
  \multicolumn{2}{c|}{\cellcolor[HTML]{FBE2D5}Dynamic} &
  \cellcolor[HTML]{FBE2D5}UAS &
  \cellcolor[HTML]{FBE2D5}2 &
  \multicolumn{1}{c|}{\cellcolor[HTML]{FBE2D5}0} &
  \multicolumn{1}{c|}{\cellcolor[HTML]{FBE2D5}1} &
  \multicolumn{1}{c|}{\cellcolor[HTML]{FBE2D5}0} &
  \multicolumn{1}{c|}{\cellcolor[HTML]{FBE2D5}0} &
  \cellcolor[HTML]{FBE2D5}2 &
  \multicolumn{1}{c|}{\cellcolor[HTML]{FBE2D5}1} &
  \cellcolor[HTML]{FBE2D5}2 \\ \cline{3-13} 
\multicolumn{1}{|c|}{\multirow{-10}{*}{\textbf{Temporal}}} &
  \multicolumn{1}{c|}{\multirow{-3}{*}{\cellcolor[HTML]{FBE2D5}Local}} &
  \multicolumn{3}{c|}{\cellcolor[HTML]{F7C7AC}\textbf{Total}} &
  \cellcolor[HTML]{F7C7AC}\textbf{4} &
  \multicolumn{1}{c|}{\cellcolor[HTML]{F7C7AC}\textbf{0}} &
  \multicolumn{1}{c|}{\cellcolor[HTML]{F7C7AC}\textbf{2}} &
  \multicolumn{1}{c|}{\cellcolor[HTML]{F7C7AC}\textbf{0}} &
  \multicolumn{1}{c|}{\cellcolor[HTML]{F7C7AC}\textbf{0}} &
  \cellcolor[HTML]{F7C7AC}\textbf{4} &
  \multicolumn{1}{c|}{\cellcolor[HTML]{F7C7AC}\textbf{2}} &
  \cellcolor[HTML]{F7C7AC}\textbf{4} \\ \hline
\multicolumn{5}{|c|}{\textbf{Cumulative Total}} &
  \textbf{100} &
  \multicolumn{1}{c|}{\textbf{8}} &
  \multicolumn{1}{c|}{\textbf{36}} &
  \multicolumn{1}{c|}{\textbf{16}} &
  \multicolumn{1}{c|}{\textbf{28}} &
  \textbf{64} &
  \multicolumn{1}{c|}{\textbf{48}} &
  \textbf{91} \\ \hline
\multicolumn{5}{|c|}{\textbf{Coverage}} &
  \multicolumn{1}{l|}{\textbf{}} &
  \multicolumn{1}{c|}{\textbf{8.00\%}} &
  \multicolumn{1}{c|}{\textbf{36.00\%}} &
  \multicolumn{1}{c|}{\textbf{16.00\%}} &
  \multicolumn{1}{c|}{\textbf{28.00\%}} &
  \textbf{64.00\%} &
  \multicolumn{1}{c|}{\textbf{48.00\%}} &
  \textbf{91.00\%} \\ \hline
\end{tabular}%
}

\small{BO: Buffer Overflow, RW: Arbitrary Read/Write, UAF: Use After Free, UAS: Use After Scope, IF: Invalid Free, DF: Double Free, \\
$^a$ Native CUDA execution without any protection. $^b$ The results are estimated based on the description of the paper.}
\vspace{-0.1in}
\end{table*}
\subsection{Coverage Evaluation}

\subsubsection{Spatial Memory Safety Analysis}
Table~\ref{tab:detailed_coverage} shows the detailed coverage results.

\niparagraph{Global Memory:} We used 16 tests to evaluate the error detection coverage in global memory. We found that baseline GPU cannot detect any global memory OOB accesses; compute sanitizer and GMOD can detect tests that use buffer-overflow but fail to detect tests that use OOB read/writes because they use tripwires for error detection. GPUShield and CuCatch can detect all tests because they use a base \& bound table to keep track of the exact bounds of the buffers. When we translate a GPU program to a CPU program, all the global memory buffers are mapped to heap buffers. On translated programs, Address Sanitizer behaves similar to Compute Sanitizer due to tripwire implementation, whereas no-Fat detects all test cases. On the CPU side, Address Sanitizer can detect tests that use buffer overflows but fails to detect OOB read/write errors. No-Fat can detect all kinds of errors in global memory. No-Fat can detect all spatial memory errors as it tracks exact base and sizes of the allocated objects.

\niparagraph{Local Memory:}
We used 28 test cases to evaluate error detection coverage in local memory. We found that baseline architecture fails to detect all buffer-overflow and OOB read/write errors. Compute sanitizer and GPUShield only detected errors that go beyond local memory bounds. GMOD fails to detect any local memory errors because it only guards global memory. CuCatch fails to detect accesses that originate from statically allocated buffers, either within or across frames, because of a lack of buffer bounds information at the compiler backend. Address sanitizer fails to detect any OOB read/write errors as it relies on tripwires. Finally, No-Fat can detect all errors in local memory.

\niparagraph{Shared Memory:}
We used 12 test cases to evaluate error detection coverage in shared memory. Baseline, GMOD, and GPUShield fail to detect any shared memory test cases because they do not guard shared memory allocations. Compute Sanitizer can only detect errors that go beyond shared memory bounds. CuCatch, AddressSanitizer, and No-Fat are unable to detect violations within dynamically allocated shared buffers, as these buffers are treated as a single contiguous allocation.

\niparagraph{Intra-allocation errors:} They
 refer to access violations that happen within an object (e.g., struct). No tools  can detect these errors except ASan, which has experimental support with  Intel MPX~\cite{intex_mpx}. This makes \mname combined with Address Sanitizer (\mname+ASan) the only technique capable of detecting intra-allocation errors in CUDA programs.

\subsubsection{Temporal Memory Safety Analysis}
\mbox{}\\ 
\niparagraph{Global memory:} We used 12 test cases to evaluate UAF, IF, and DF errors in global memory. All GPU tools detected IF and DF errors, as these are supported by the underlying GPU memory allocator. The baseline allocator cannot detect UAF errors. GMOD and GPUShield fail to detect UAF errors since they focus only on spatial safety. Compute Sanitizer, CuCatch, and Address Sanitizer detect immediate UAF errors, and CuCatch can also probabilistically detect delayed UAF errors. CPU tools can only detect one IF case involving a non-base pointer in a ~\texttt{free()} call. Cases where a \texttt{free()} call is made with a pointer from a different GPU memory allocator are not detected because \mname does not differentiate between different GPU memory allocators.

\niparagraph{Local Memory:} We used 4 tests to detect UAS violations in local memory. We found that the baseline GPU cannot detect UAS tests. Since GMOD and GPUShield don't focus on temporal safety, they fail to detect all UAS test cases. Compute Sanitizer and Address Sanitizer can only detect immediate UAS violations, while CuCatch can detect immediate UAS errors and probabilistically detect delayed UAS errors. Finally, No-Fat can detect both immediate and delayed UAS errors.

\subsection{Fuzzing with \mname and ASan}

Table~\ref{tab:fuzzing_results} summarizes crashes and hangs from our fuzzing campaigns. The \textit{total finds} column reports unique issues identified by \mname after AFLTriage~\cite{afltriage} postprocessing, representing cases developers must address.

\begin{table}[htb]
\caption{Fuzzing Results with \mname + AddressSanitizer}
\label{tab:fuzzing_results}
\centering
\resizebox{\columnwidth}{!}{%
\begin{tabular}{cllllllll}
\hline
\multirow{2}{*}{\textbf{Benchmark}} &
  \multicolumn{3}{c}{\textbf{Crashes}} &
  \multicolumn{1}{c}{\multirow{2}{*}{\textbf{Hangs}}} &
  \multicolumn{1}{c}{\multirow{2}{*}{\textbf{Total Finds}}} &
  \multicolumn{1}{c}{\multirow{2}{*}{\textbf{Cycles}}} &
  \multicolumn{1}{c}{\multirow{2}{*}{\textbf{Exec/Sec}}} &
  \multicolumn{1}{c}{\multirow{2}{*}{\textbf{Max Depth}}} \\ \cline{2-4}
 &
  \textbf{Kernel} &
  \textbf{Host} &
  \textbf{Total} &
  \multicolumn{1}{c}{} &
  \multicolumn{1}{c}{} &
  \multicolumn{1}{c}{} &
  \multicolumn{1}{c}{} &
  \multicolumn{1}{c}{} \\ \hline
    aes~\cite{heteromark}             & 0 & 3  & 3  & 0  & 3  & 17  & 33.68 & 8  \\
    bs~\cite{heteromark}               & 1 & 2  & 3  & 4  & 7  & 225 & 34.97 & 5  \\
    ep~\cite{heteromark}               & 0 & 4  & 4  & 8  & 12 & 9   & 1.81  & 4  \\
    fir~\cite{heteromark}              & 0 & 2  & 2  & 5  & 7  & 6   & 1.43  & 7  \\ 
    ga~\cite{heteromark}               & 0 & 11 & 11 & 15 & 26 & 2   & 17.45 & 24 \\
    hist~\cite{heteromark}             & 0 & 2  & 2  & 5  & 7  & 122 & 32.86 & 8  \\
    kmeans~\cite{heteromark}           & 0 & 0  & 0  & 0  & 0  & 4   & 30.55 & 14 \\
    pr~\cite{heteromark}               & 4 & 3  & 7  & 4  & 11 & 17  & 1.33  & 9  \\ \hline
    vecadd~\cite{cuda_sample}          & 0 & 2  & 2  & 4  & 6  & 51  & 7.4   & 5  \\
    matmul~\cite{cuda_sample}           & 2 & 2  & 4  & 1  & 5  & 238 & 80.55 & 5  \\ \hline
    bfs~\cite{rodinia}             & 4 & 2  & 6  & 7  & 13 & 11  & 7.9   & 6  \\
    cfd~\cite{rodinia}              & 3 & 3  & 6  & 7  & 13 & 0   & 0.32  & 4  \\
    gaussian~\cite{rodinia}         & 0 & 2  & 2  & 5  & 7  & 0   & 1.81  & 2  \\
    lud~\cite{rodinia}              & 1 & 2  & 3  & 2  & 5  & 1   & 0.71  & 9  \\ \hline
    \textbf{Total}  & 15& 40 & 55 & 67 & 122&     &       &    \\ \hline
\end{tabular}%
}
\end{table}

We conducted fuzzing campaigns on a set of CUDA benchmarks using \mname with Address Sanitizer. This enabled the detection of memory access violations in both the host and kernel components of the CUDA programs. During the campaign, we uncovered several crashes, which were primarily attributed to malformed input files. These crashes were caused by structural inconsistencies or the presence of invalid or corrupted data in the input files (found in~\texttt{aes}). Some input files included allocation sizes and other setup parameters, where corruption triggered integer overflows. This, in turn, led to requests for excessively large allocation sizes from the memory allocator, ultimately resulting in crashes. We found this issue in \texttt{vecadd}, \texttt{matmul}, \texttt{bfs}, \texttt{cfd}, \texttt{gaussian}, and \texttt{lud} benchmarks. In the kernel code, we identified several global memory out-of-bounds read and write accesses in the \texttt{matmul}, \texttt{bfs}, \texttt{cfd}, and \texttt{lud} benchmarks. Similarly, in the host code, we observed multiple out-of-bounds read and write accesses in the \texttt{aes}, \texttt{ga}, \texttt{pr}, and \texttt{bfs} benchmarks. These errors manifested as heap-buffer overflows in the translated program. Additionally, we discovered cases of direct null pointer dereferences in \texttt{cfd} and instances where null pointers were passed to the \texttt{memcpy()} function in \texttt{ga} and \texttt{ep}. Several benchmarks, like \texttt{bs}, \texttt{ep}, \texttt{fir}, \texttt{hist} also reported unhandled exceptions thrown by the argument parser component of these benchmarks. 

\mname runtime dynamically calculates the number of tasks to execute. Interestingly, from a crash due to malformed input, we discovered that this calculation logic was not safeguarded against corner cases, such as division by zero when the input included a value of zero. This issue was observed across multiple benchmarks, including \texttt{vecadd}, \texttt{matmul}, \texttt{aes}, \texttt{pr}, \texttt{cfd}, \texttt{fir}, and \texttt{lud}. Notably, this allowed us to identify a bug within the \mname runtime.

\subsubsection{Security Implications of the found bugs}
While many crashes arise from malformed inputs, their root cause like buffer-overflows, OOB accesses and allocator misuses are security critical. In real-world deployments, such bugs could be exploited by attackers by supplying crafted inputs to GPU accelerated services. For instance, OOB writes in kernels (\texttt{matmul}, \texttt{bfs}, \texttt{cfd}, \texttt{lud}) could corrupt adjacent data structures and potentially enable control-flow hijacking on GPU~\cite{mind_contol_attack_2021, fun_and_profit_2024}. Heap-buffer overflows and null pointer dereferences in host code (\texttt{aes}, \texttt{ga}, \texttt{pr}, \texttt{bfs}) creates classic CPU-side attack vectors, allowing bugs in CUDA program to escalate to system-level compromises. Furthermore, Crashes due to invalid allocation or divide by zero conditions can be weaponized to repeatedly crash services leading to denial-of service attacks. Collectively, these findings demonstrate that GPU memory errors are not marely correctness bugs but exploitable vulnerabilities with consequences for confidentiality, integrity and availability of GPU applications. By exposing such flaws systematically, \mname highlights the urgent need for memory-safety mechanisms in GPU programming.

\subsection{Performance Evaluation}
\label{sec:performance_evaluation}

\begin{figure*}[htbp]
    \begin{center}
        \mbox{ 
        \hspace{-1.0ex}
        \subfigure[Throughput Improvement brought by PREX]
            {
            \label{fig:prex_improvement}
            \includegraphics[width=0.31\textwidth]{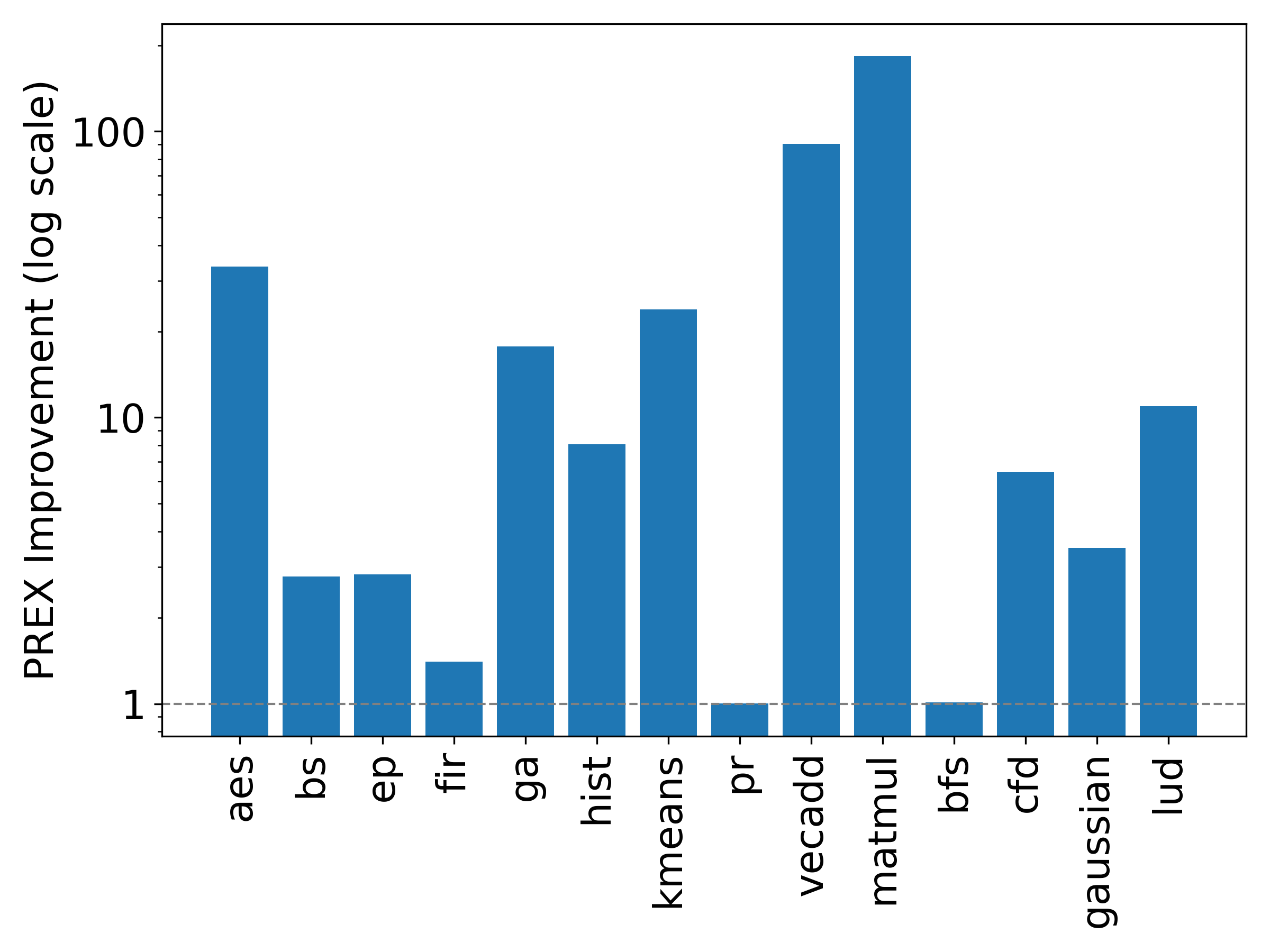}
            }
        \hspace{-1.0ex}
        \subfigure[Throughput Improvement brought by AXIPrune]
            {
            \label{fig:axiprune_improvement}
            \includegraphics[width=0.31\textwidth]{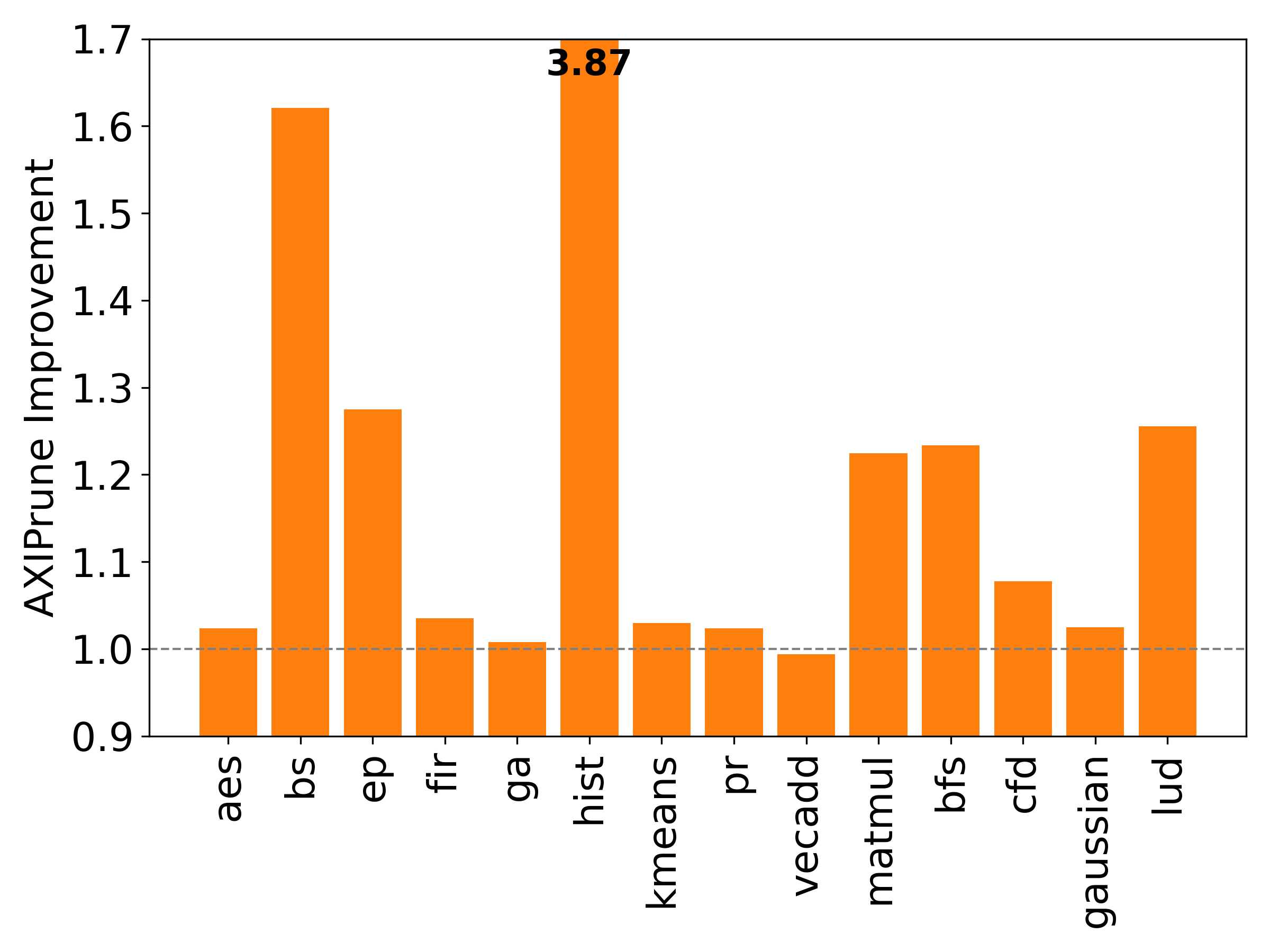}
            }
        \hspace{-1.0ex}
        \subfigure[Executed Instructions w/o and w/ AXIPrune]
            {
            \label{fig:axiprune_instr_count}
            \includegraphics[width=0.31\textwidth]{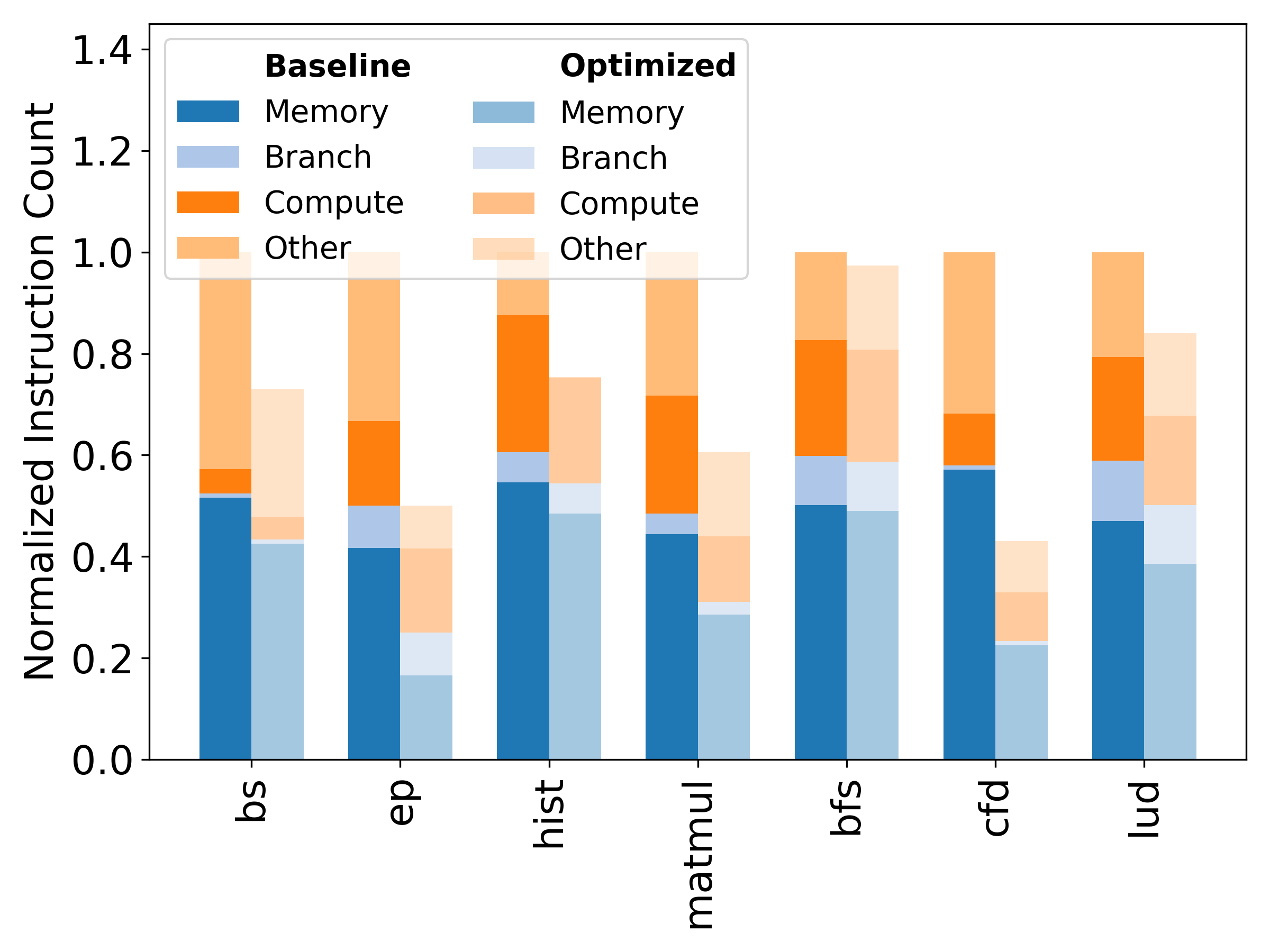}
            }
        }
           \vspace{-2.0ex}
        \caption{Performance Analysis of \mname Optimizations}
        \label{fig:perf_analysis} 
    \end{center}
\end{figure*}

The fuzzing process involves executing the same program with a large number of distinct inputs to detect potential bugs. As throughput is the primary metric of execution efficiency, we use it to evaluate the effectiveness of the PREX and AXIPrune optimizations.

\subsubsection{PREX}

As demonstrated in Figure~\ref{fig:prex_improvement}, PREX significantly improves throughput in 12 out of the 14 benchmarks. On average, PREX achieves a 32× speedup compared to the baseline. For the \texttt{pr} and \texttt{bfs} benchmarks, the compiler analysis detects that the GPU kernels contain indirect memory accesses, which do not satisfy the requirements of affine access. As a result, these kernels are not executed partially and exhibit the same throughput as the baseline.

PREX achieves the most significant improvements on the \texttt{aes}, \texttt{vecadd}, and \texttt{matmul} benchmarks. All of these benchmarks contain a large number of threads and blocks in the original GPU kernels, and all memory accesses are affine with respect to the thread and block indices. By applying the PREX optimization, \mname only needs to execute the first and last threads and blocks. By significantly reducing the number of blocks and threads that need to be executed, PREX achieves throughput improvements of up to 183$\times$.

\subsubsection{AXIPrune}

The throughput improvement achieved by applying AXIPrune is shown in Figure~\ref{fig:axiprune_improvement}. These results are based on code already optimized with PREX. AXIPrune provides substantial gains for benchmarks with heavy computational workloads. For example, the \texttt{ep} benchmark contains an exponential function call that is computationally expensive yet unrelated to memory access or bug detection. AXIPrune eliminates this call, resulting in a 27\% throughput improvement.

AXIPrune also optimizes kernels with complex workflows. For example, the \texttt{hist} benchmark includes a barrier instruction to enforce memory consistency. After being transformed into a CPU program, the barrier introduces complex control flow by generating separate for-loops. In the \texttt{hist} benchmark, memory consistency affects only the numerical output values and does not influence memory access indices. Therefore, AXIPrune detects barriers that are unrelated to memory access indices and eliminates them, resulting in simpler and more efficient CPU programs.

To further demonstrate the effect of the AXIPrune optimization, we profile the executed instructions and present the results in Figure~\ref{fig:axiprune_instr_count}. AXIPrune eliminates instructions that are not related to memory access in the original GPU programs. As a result, it reduces the latency of executing GPU programs on CPUs, which in turn increases the throughput of \mname within a given timeframe.

Note that The reduction in memory access instructions shown in Figure~\ref{fig:axiprune_instr_count} does not contradict our claim that AXIPrune preserves all original GPU memory accesses. In CPU programs without AXIPrune, two types of accesses exist: (1) original accesses from the GPU program and (2) compiler-induced accesses introduced during GPU-to-CPU translation. AXIPrune preserves all original accesses—ensuring that memory bugs from the GPU program remain visible—while eliminating only compiler-induced accesses. For example, in Listing~\ref{lst:pact_transformed_gpu_prog}, lines 9 and 10, the access to \texttt{id[tid]} is an auxiliary memory access introduced by the transformation process and does not exist in the original GPU program (Listing~\ref{lst:pact_orig_gpu_prog}). These auxiliary accesses are unrelated to the memory bugs in the original GPU program and can be safely eliminated by AXIPrune to improve \mname throughput.
In general, for all programs optimized by AXIPrune, based on the improvements already brought by PREX, AXIPrune can further increase the throughput by 33\% on average.

%% file: sources/09_related.tex
\section{Related Works}
\label{sec:related_works}
\paragraph{Fuzzing Techniques}
Software fuzzing has been extensively studied in the context of CPU programs, with notable efforts including AFL/AFL++~\cite{afl, aflpp}, Libfuzzer~\cite{libfuzzer}, Hongfuzz~\cite{honggfuzz} among others~\cite{ref_fuzzgen,ref_fuzzguard,ref_fudge,ref_fuzz4all}. AFL~\cite{afl} is a general-purpose fuzzer that uses a genetic-algorithm-driven coverage-guided fuzzing technique to discover vulnerabilities. AFL++~\cite{aflpp} extends this by providing a flexible framework for combining diverse fuzzing methods. CLFuzz~\cite{ref_clfuzz} proposes a semantic-aware fuzzing method with adaptive input generation and logical cross-checks, enhancing vulnerability detection in cryptographic algorithms. However, all of these approaches are designed for CPU-based applications. There has been little work on GPU fuzzing. One notable exception is the work by Peng~\etal~\cite{acm2020_opencl_fuzzer}, who developed an automated test generation method for OpenCL kernels that integrates mutation-based fuzzing with selective constraint solving.

\paragraph{GPU Memory Safety}
There have been many recent studies~\cite{ref_gpu_mem_safety,ref_cucatch,ref_gpushield} on GPU memory safety due to the unique security challenges posed by GPU architectures, such as their distinct memory hierarchies and concurrent execution models. Guo et al.\cite{ref_gpu_mem_safety} propose a comprehensive analysis of buffer overflow vulnerabilities in modern GPUs, showing that these issues can lead to code injection and return-oriented programming attacks, posing significant security risks for GPU applications. 
GPUShield~\cite{ref_gpushield} proposes a hardware-based bounds-checking mechanism to ensure spatial memory safety in GPU memory spaces. 
CuCatch~\cite{ref_cucatch} proposes an efficient debugging tool for CUDA applications that detects spatial and temporal memory violations with low performance overhead, using optimized compiler instrumentation and driver support.
IMT~\cite{sullivan2023implicit} introduces a memory tagging mechanism that eliminates the need for additional memory storage by utilizing redundant bits in ECC. However, this approach relies on a probabilistic method, making it susceptible to brute-force attacks.

\paragraph{GPU to CPU migration tools} Existing GPU-to-CPU migration techniques can be broadly categorized based on the level at which the transformation is applied: Source-to-source translators~\cite{cumulus, hipify, mcuda} convert high-level CUDA code into equivalent CPU-executable source code (e.g., C++, HIP, OpenMP, etc.). IR-level translators~\cite{cupbop_todaes} operate on intermediate representations such as LLVM IR by performing IR-level transformations to map GPU execution models to CPUs. Lastly, execution emulators~\cite{ocelot} interpret or simulate compiled GPU binaries directly on the CPU. CuPBoP~\cite{cupbop_todaes} is an open-source framework that provides an IR-level translator. However, CuPBoP lacks support for systematic exploration of memory safety vulnerabilities through dynamic analysis techniques because it does not integrate seamlessly with fuzzing tools like AFL. Additionally, CuPBoP does not preserve debugging symbols across its transformations, limiting its suitability for debugging and fuzz-testing workflows. While CuPBoP prioritizes execution correctness with conservative runtime strategies, it lacks any optimizations aimed at enhancing fuzzing throughput and performance. 

\mname directly addresses these shortcomings by carefully preserving symbols throughout the transformation pipeline, enabling meaningful crash analysis and debugging during fuzz testing. Additionally, \mname introduces targeted optimizations, including partial representative execution and pruning methods, significantly improving fuzzing efficiency by reducing unnecessary execution overhead. These enhancements allow \mname to integrate seamlessly with established CPU fuzzing tools like AFL and AddressSanitizer. As a result, \mname effectively enables the discovery of GPU-specific memory safety vulnerabilities, a capability currently absent in existing tools such as CuPBoP.

%% file: sources/10_conclusion.tex
\section{Conclusions and Future Work}
\label{sec:conclusion}
This work presents \mname, the first proposal to apply fuzzing to GPU programs by utilizing GPU-to-CPU code translation to detect memory safety bugs. To demonstrate its effectiveness, we developed GMSBenchmark and evaluated 14 GPU benchmarks, uncovering several previously unknown memory safety violations. Furthermore, our two compile-time pruning techniques significantly accelerate fuzzing by reducing execution time. We believe this work lays the foundation for future research in GPU fuzzing and advancing GPU security.